\input harvmac 
\input epsf.tex
\def\IN{\relax{\rm I\kern-.18em N}} 
\def\IR{
\relax{\rm I\kern-.18em R}} \font\cmss=cmss10 
\font\cmsss=cmss10 at 7pt \def\IZ{\relax\ifmmode\mathchoice 
{\hbox{\cmss Z\kern-.4em Z}}{\hbox{\cmss Z\kern-.4em Z}} 
{\lower.9pt\hbox{\cmsss Z\kern-.4em Z}} {\lower1.2pt\hbox{
\cmsss Z\kern-.4em Z}}
\else{\cmss Z\kern-.4em Z}\fi} 

\overfullrule=0mm

\newcount\figno \figno=0
\newcount\figtotno      
\figtotno=0
\newdimen\captionindent 
\captionindent=1cm 
 
\newcount\figno
\figno=0
\def\fig#1#2#3{ \par\begingroup\parindent=0pt
\leftskip=1cm\rightskip=1cm\parindent =0pt 
\baselineskip=11pt
\global\advance\figno by 1
\midinsert
\epsfxsize=#3
\centerline{\epsfbox{#2}}
\vskip 12pt
{\bf Fig. \the\figno:} #1\par
\endinsert\endgroup\par
}
\def\figlabel#1{\xdef#1{\the\figno}} 
\def\encadremath#1{\vbox{\hrule\hbox{\vrule\kern8pt 
\vbox{\kern8pt \hbox{$\displaystyle #1$}\kern8pt} 
\kern8pt\vrule}\hrule}} \def\enca#1{\vbox{\hrule\hbox{
\vrule\kern8pt\vbox{\kern8pt \hbox{$\displaystyle #1$}
\kern8pt} \kern8pt\vrule}\hrule}}

\def\IR{\relax{\rm I\kern-.18em R}}
\font\cmss=cmss10 \font\cmsss=cmss10 at 7pt 
\def\IZ{\relax\ifmmode\mathchoice
{\hbox{\cmss Z\kern-.4em Z}}{\hbox{\cmss Z\kern-.4em Z}} 
{\lower.9pt\hbox{\cmsss Z\kern-.4em Z}}
{\lower1.2pt\hbox{\cmsss Z\kern-.4em Z}} 
\else{\cmss Z\kern-.4em Z}\fi} \def\buildrel#1\under#2{ 
\mathrel{\mathop{\kern0pt #2}\limits_{#1}}}

\Title{UNC-CH-MATH-97/15, DTP-97/57 and T97/133}
{{\vbox {
\centerline{Coloring Random Triangulations}}}}
\medskip
\centerline{P. Di Francesco\footnote*{e-mail: philippe@math.unc.edu},}
\smallskip
\centerline{\it Department of Mathematics,} 
\centerline{\it University of North Carolina at Chapel Hill,} 
\centerline{\it  CHAPEL HILL, N.C. 27599-3250, U.S.A.} 
\medskip
\centerline{B. 
Eynard\footnote{${ }^{\#} $}{e-mail: Bertrand.Eynard@durham.ac.uk}} 
\smallskip
\centerline{\it Department of Mathematical Sciences,}
\centerline{\it University of Durham, Science Labs,}
\centerline{\it South Road, DURHAM DH1 3HP, U.K.}
\medskip
\centerline{E. Guitter\footnote{${ }^{\$ } $}{e-mail: guitter@spht.saclay.cea.fr}}
\smallskip
\centerline{\it Service de Physique Th\'eorique,} 
\centerline{\it C.E.A. Saclay,}
\centerline{ \it F-91191 Gif sur Yvette Cedex, France} 
\vskip .5in
\noindent 

We introduce and solve a two-matrix model for the tri-coloring problem
of the vertices of a random triangulation. We present three different
solutions: (i) by orthogonal polynomial techniques (ii) by use
of a discrete Hirota bilinear equation (iii) by direct expansion.
The model is found to lie in the universality class of pure
two-dimensional quantum gravity, despite the non-polynomiality of its
potential.
\vfill\noindent
P.A.C.S. numbers: 05.20.y, 04.60.Nc --
\tenpoint\supereject\global\hsize=\hsbody
\footline={\hss\tenrm\folio\hss}

\nref\BAX{R. Baxter, {\it Exactly Solved Models of Statistical
Mechanics}, Oxford (1982).}
\nref\BCOL{R. Baxter, J. Math. Phys. {\bf 11} (1970) 784;
J. Phys. {\bf A19} (1986) 2821.}
\nref\DG{P. Di Francesco and E. Guitter, 
Europhys. Lett. {\bf 26} (1994) 455, cond-mat/9406041.}
\nref\CMM{G. Cicuta, L. Molinari and E. Montaldi, Phys. Lett.
{\bf B306} (1993) 245.}
\nref\CK{L. Chekhov and C. Kristjansen, Nucl. Phys. {\bf B479}
(1996) 683, hep-th/9605013.}
\nref\KE{C. Kristjansen and B. Eynard, Durham preprint DTP-97/55 , 
cond-mat/9710199.}
\nref\DGZ{P. Di Francesco, P. Ginsparg
and J. Zinn--Justin, {\it 2D Gravity and Random Matrices},
Physics Reports {\bf 254} (1995) 1, hep-th/9306153.}
\nref\Penner{R.C. Penner, Comm. Math. Phys. {\bf 113} (1987) 299;
J. Diff. Geom. {\bf 27} (1988) 35.}
\nref\Mak{Yu. Makeenko, Phys. Lett. {\bf B314} (1993) 197, hep-th/9306043.}
\nref\KM{V.A. Kazakov and A.A. Migdal, Nucl. Phys {\bf B397} (1993) 214.}
\nref\IZ{C. Itzykson and J.-B. Zuber, J. Math. Phys. {\bf 21} (1980)
411.}
\nref\KAZ{V. Kazakov, private communication.}
\nref\WIG{I. Krichever, O. Lipan, P. Wiegmann and A. Zabrodin,
Comm. Math. Phys. {\bf 188} (1997) 267, hep-th/9604080.}
\nref\TUT{W. Tutte, 
Canad. Jour. of Math. {\bf 15} (1963) 249.}
\nref\DGG{P. Di Francesco, O. Golinelli
and E. Guitter, 
Mathematical and Computer
Modeling {\bf 144} (1997), hep-th/9506030.}
\nref\DGGB{P. Di Francesco, O. Golinelli and E. Guitter,
Comm. Math. Phys.  {\bf 186} (1997) 1, hep-th/9602025.}
\nref\SM{this problem is discussed in the mathematical entertainments
section of the Mathematical Intelligencer Volume 19 number 4 (1997) 48.}

\newsec{Introduction}

Various graph coloring problems have been considered in the last two 
decades in relation with statistical mechanics. On regular lattices, 
many of these have proved to be integrable and exactly solvable \BAX.
Among these is the classical tri-coloring problem of the bonds of the 
triangular lattice, solved by Baxter in \BCOL: in how many ways can 
one color the lattice bonds with three colors so that any triangle 
has its three edges of a different color. Recently a remarkable 
connection was made between this problem and that of folding
of the triangular lattice \DG, counting the different ways to fold 
the lattice onto itself along its bonds, a toy model for the crumpling
of two-dimensional polymerized membranes.
A natural extension of these problems consists in their generalization
to random triangulations. This may be viewed as a coupling of the model to 
two-dimensional quantum gravity, or also as a way to describe fluid membranes. 
The bond tri-coloring of random triangulations has first been addressed 
in \CMM\ and solved in \CK\ for a restricted model where the colors 
always occur in the same cyclic order for all triangles. The complete 
bond tri-coloring problem without restriction was solved very recently 
in \KE . The connection to the folding of random triangulations requires 
unfortunately an additional constraint: the number of triangles around 
each vertex must be even. The folding of random triangulations remains 
an open problem.

In this paper, we address a different kind of (bi- or tri-)coloring 
problems, where we now color the vertices or the faces of some 
random graphs, representing tessellated surfaces of arbitrary genus. 
Starting from the vertex bi-coloring of arbitrary graphs, our main result 
will eventually be rephrased as solving the tri-coloring problem of the 
vertices of random triangulations of surfaces. 

Our starting point is the computation of the following integral over 
two Hermitian matrices
$M$ and $R$ of size $N\times N$ (see \DGZ\ for a review)
\eqn\pfunc{ Z(p,q,g;N)~=~\int dM dR e^{-N\, {\rm Tr}\, V(M,R;p,q,g)} }  
with the non-polynomial potential
\eqn\pot{ V(M,R;p,q,g)~=~ 
p{\rm Log}(1-M) +q 
{\rm Log}(1-R) +g MR }
and with respect to the Haar measure over Hermitian matrices $H$
\eqn\haar{ dH~=~ c_N(g)\, \prod_{i=1}^N dH_{ii} \, \prod_{1\leq i<j\leq N}
d{\rm Re}H_{ij}\, d{\rm Im}H_{ij} }
normalized in such a way that $Z(0,0,g;N)=1$ for all $g$ and $N$.
The parameters $p,q,g$ are {\it a priori} any real numbers.
This model, with its logarithmic potential, can be viewed as a two-matrix 
generalization of the one-matrix Penner model \Penner.
For $p=q$, it has already been studied in \Mak\ in connection with
the Kazakov-Migdal model \KM. 

Upon setting $g=1/t$ and rescaling $M\to \sqrt{t} M$, $R \to \sqrt{t} R$
in the integral \pfunc, we may rewrite \pfunc\ in the same form, but
with the potential
\eqn\newpot{ W(M,R;p,q,1/t)~=~
p {\rm Log}(1-\sqrt{t} M) +q 
{\rm Log}(1-\sqrt{t}R) + MR }
and with a different normalization of the measure, still ensuring
that $Z(0,0,1/t;N)=1$. Note that this last condition is equivalent to
demanding that $\lim_{t\to 0} Z(p,q,1/t;N)=1$.
If we think of $t$ as a small parameter, we
may at least formally expand the integral $Z$ in powers of $t$.
This expansion is expressible as a sum over Feynmann (fat)graphs,
made of double-lines oriented in opposite directions and carrying
a matrix index $i\in \{1,2,...,N\}$,
involving two types of vertices ($M$-vertices weighed by $Np$
and $R$-vertices weighed by $Nq$, irrespectively of the valency) 
connected by one type of propagator
($\langle MR\rangle $ propagator, weighed by $t/N$). 
As usual in matrix integrals, each graph also receives a contribution
$N$ per oriented loop, from the summation over the running matrix indices.
Hence we finally get the expansion
\eqn\graxp{ Z(p,q,1/t;N)~=~\sum_{{\rm fatgraphs}\ \Gamma}
{t^{E(\Gamma)}\, N^{\chi(\Gamma)} \,  p^{V_p} 
q^{V_q} \over |{\rm Aut}(\Gamma)|} }
where $E(\Gamma)$ denotes the number of edges of $\Gamma$, $\chi(\Gamma)$ 
its Euler
characteristic, $V_p$, $V_q$ 
its numbers of $M$ and $R$-vertices respectively,  and $|$Aut$(\Gamma)|$
the order of its symmetry group. The sum may be restricted to connected 
fatgraphs by considering Log$\, Z$ instead.

If we choose to color the {\it vertices} of the fatgraphs with two different 
colors, say white for the $M$-vertices and black for the $R$-vertices, 
the occurrence of only mixed propagators $\langle MR\rangle$ implies that 
the graph is bi-colored, namely two adjacent vertices have different colors.
The white (resp. black) faces receive the weight $p$ (resp. $q$). 
We can also consider \graxp\ as an expansion over the dual fatgraphs, 
with now black and white alternating {\it faces}.
So we may finally view Log$\, Z$ as the partition
function for connected bi-colored fatgraphs of arbitrary genus.

When compared to the class of exactly computable multi-matrix integrals,
\pfunc\ is readily seen to be reducible to an integral over the eigenvalues
of the matrices $M$ and $R$, by use of the Itzykson-Zuber integral formula
\IZ. However, we expect the non-polynomiality of the potential \pot\ to
translate into some interesting non-locality of the model, which makes
its solution slightly more subtle than that of the polynomial case.

\par
\medskip
The paper is organized as follows. 
The model \pfunc\ is solved in Sect.2, using standard orthogonal
polynomial techniques. The solution involves the computation of partition 
functions $Z_n$ with the same definition as \pfunc, but where the 
integration extends over $n\times n$ Hermitian matrices, for all
values of $n$ between $1$ and $N$.
The varying size $n$ of the matrices introduces a new scaling variable 
$z=n/N$ in the problem. Remarkably, the solution exhibits an explicit 
$(p,q,z)$ symmetry relating this size scaling variable $z$ to the coupling
constants $p$ and $q$ of the model.
This symmetry is explained by identifying our problem with that of 
tri-coloring the vertices of random triangulations, with the
third color (say grey) receiving precisely the weight factor $z$.
The solution is further explored by taking the large
$N$ limit, in which only planar (genus zero) diagrams contribute.
The result is a non-linear partial differential equation for the
free energy, involving derivatives wrt $t$ and $z$. 
To solve it, we first derive initial data when $z\to 0$,
in which case the free energy may be explicitly computed (the
detailed derivation is done in Appendix A).
In Sect.3, we present an alternative solution of our model, by
first writing the integral \pfunc\ as a determinant, and then
deriving a (discrete Hirota) bilinear equation which determines
the partition function completely. In the large $N$ limit,
this takes the form of yet another non-linear partial differential
equation for the free energy, involving derivatives wrt $p$, $q$,
$z$ and $t$.
Combining the results of Sects.2 and 3., we are able to
compute explicitly the free energy in the case $p=q=z$.
In Sect.4, we follow a third route to express the partition
function $Z_n$ of our model as an explicit multiple sum
over strictly increasing sequences of integers. We further
extract the large $N$ behavior of these sums by applying the
saddle-point approximation, which yields yet another expression
for the planar free energy. The latter permits to investigate the
critical properties of the model. We find a first order critical 
point whenever $p$, $q$ and $z$ have the same sign, with a generic 
singularity ${\rm Log} \,Z\sim(t_*-t)^{5/2}$, characteristic of the
universality class of pure two-dimensional quantum gravity.
The all-genus double-scaling limit of the free energy is also derived,
and shown to be governed by the Painlev\'e I differential equation.
We conclude in Sect.5 with a discussion on the connection of our
model to folding problems and with some remarks on the emergence 
of discrete Hirota bilinear equations in multi-matrix models.

\par
\newsec{Orthogonal Polynomial Solution}
\par
\subsec{Preliminaries}
\par
As mentioned above, we first reduce the integral \pfunc\ to an integral
over the eigenvalues of $M$ and $R$, denoted by $m_i$ and $r_i$ respectively.
We simply perform the change of variables $(M;R)\to (m,U; r,V)$, with
$M=UmU^\dagger$ and $R=VrV^\dagger$, $U,V$ two unitary matrices, and
$m=$diag$(m_1,...,m_N)$, $r=$diag$(r_1,...,r_N)$.  
The Jacobian of the change of variables is simply $\Delta(m)^2\Delta(r)^2$,
where $\Delta(m)=\prod_{i<j} (m_i-m_j)$ denotes the Vandermonde determinant
of the matrix $m$.
The integral over the unitary matrices 
$U$ and $V$ only involves the cross-term $-Ng{\rm Tr}(MR)$ 
in the potential and reads
\eqn\iniz{ \int dU dV e^{-Ng \,{\rm Tr}(MR)} ~=~ d_N(g){\det \big[
e^{-Ng m_i r_j}
\big]_{1\leq i,j \leq N} \over \Delta(m) \Delta(r) } }
by direct use of the Itzykson-Zuber integral formula. The precise value of
the normalization constant $d_N(g)$ is not important here.
Substituting this into the integral \pfunc, and using the antisymmetry of the
Vandermonde determinant, we finally get the reduced 
integral
\eqn\redint{ Z(p,q,g;N)~=~ \int dm dr \Delta(m)\Delta(r) 
e^{-N{\rm Tr}V(m,r;p,q,g)} }
where the integral extends over the $2N$ real variables $m_i,r_i$, and
the $N$-dimensional measure over diagonal matrices
$l=$diag$(l_1,...,l_N)$
\eqn\measu{dl~=~\sqrt{e_N(g)}\, \prod_{i=1}^N dl_i} 
is normalized so as to ensure that $Z(0,0,g;N)=1$.  The normalization 
factor $e_N(g)$ will be computed later.

As in the standard orthogonal polynomial technique, we introduce two sets
of {\it monic} polynomials $p_n(x)=x^n+$lower degree, and 
${\tilde p}_n(y)=y^n+$lower degree, for $n=0,1,2,...$, which are orthogonal 
with respect to the one-dimensional measure inherited from \redint, namely
\eqn\orto{ (p_n, {\tilde p}_m)~=~
\int dx dy e^{-NV(x,y;p,q,g)} p_n(x) {\tilde p}_m(y)~=~h_n\delta_{m,n}}
for some normalization factors $h_n(p,q,g;N)$, 
entirely fixed by the orthogonality condition.  
Using the multi-linearity of the determinants, we may rewrite
\eqn\monic{\eqalign{
\Delta(m)~&=~\det\, [m_i^{j-1}]_{1\leq i,j \leq N}~=~
\det\, [p_{j-1}(m_i)]_{1\leq i,j \leq N} \cr
\Delta(r)~&=~\det\, [r_i^{j-1}]_{1\leq i,j \leq N}~=~
\det\, [{\tilde p}_{j-1}(r_i)]_{1\leq i,j \leq N}\cr}}
and expand the latter. Using the orthogonality relations, the
partition function is finally rewritten as
\eqn\parfi{\encadremath{ 
Z(p,q,g;N)~=~ N! \, e_N(g)\, \prod_{i=0}^{N-1} h_i }}

\subsec{$P$, $Q$ operators and relations}

Next we write recursion relations linking the $h_i$'s by introducing
the operators $Q$ and $\tilde Q$ of multiplication by an eigenvalue
of respectively $M$ and $R$, namely 
\eqn\defqt{(Qp_n)(x)~=~xp_n(x) \qquad 
({\tilde Q}{\tilde p}_n)(y)~=~y {\tilde p}_n(y)} 
and the operators $P$ and $\tilde P$ defined by
\eqn\pptil{  
(Pp_n)(x)~=~ (x-1) {dp_n(x) \over dx} \qquad 
({\tilde P}{\tilde p}_n)(y)~=~ (y-1) {d{\tilde p}_n(y) \over dy} }
Note that these differ slightly from the usual definitions for the 
polynomial potential case (where $P$ and $\tilde P$ are simply the 
derivatives wrt the corresponding eigenvalue), and that they satisfy
different (non-canonical) commutation relations
\eqn\commu{ [P,Q]~=~Q-1 \qquad [{\tilde P},{\tilde Q}]~=~ {\tilde Q}-1}

Equations for the $h_i$'s are derived by considering various matrix
elements of the operators $P$ and $\tilde P$.
\eqn\matop{\eqalign{ 
(Pp_n,{\tilde p}_m)~&=~-(p_n,{\tilde p}_m)
+N( (x-1)\partial_x V(x,y;p,q,g) p_n,{\tilde p}_m) \cr
&=~(Np-1)(p_n,{\tilde p}_m)+Ng( (Q-1)p_n,{\tilde Q}{\tilde p}_m)\cr}}
where we have performed an integration by parts wrt $x$. Note that
the particular definition of $P$ is ad-hoc to eliminate all
non-polynomial dependence on $Q$ in the final equation. 
Denoting by $A^\dagger$ the adjoint of an operator $A$ wrt the 
bilinear form $(f,Ag)=(A^\dagger f,g)$, we can recast \matop\
into an operator identity
\eqn\opid{ P +(1-Np) ~=~Ng {\tilde Q}^\dagger (Q-1) }
Analogously, we have
\eqn\othp{ {\tilde P} +(1-Nq)~=~NgQ^\dagger ({\tilde Q}-1)}
The operator relations \commu, \opid\  and \othp\ 
determine the $h_i$'s entirely.

Note that in our definitions, $P,Q$ and ${\tilde P},{\tilde Q}$
act on two different bases $(p_n)$ and $({\tilde p}_n)$ of the
space of polynomials of one variable.  Let us now 
write the actions of  the operators $P$, $Q$, 
${\tilde P}^\dagger$ and ${\tilde Q}^\dagger$
on the {\it same} basis $(p_n)$.
We have
\eqn\supdef{\eqalign{
(Qp_n)(x)~&=~\sum_{k\geq -1} Q_{n,k}\,  p_{n-k}(x) \cr
(Pp_n)(x)~&=~\sum_{k\geq 0} P_{n,k}\,  p_{n-k}(x) \cr
({\tilde Q}^\dagger p_n)(x)~&=~
\sum_{k \geq -1} {\tilde Q}_{n,k}\,  p_{n+k}(x) \cr
({\tilde P}^\dagger p_n)(x)~&=~
\sum_{k\geq 0} {\tilde P}_{n,k}\,  p_{n+k}(x) \cr}}
where the first terms are given by
\eqn\lead{\eqalign{ Q_{n,-1}~&=~ 1 \cr
P_{n,0}~&=~ n \cr
{\tilde Q}_{n,-1}~&=~{h_n\over h_{n-1}}\cr
{\tilde P}_{n,0}~&=~n \cr}}
The first two lines of \lead\ follow directly from the definitions
\pptil, whereas
the last two lines are easily derived from the identities 
\eqn\ideno{\eqalign{
({\tilde Q}^\dagger p_n,{\tilde p}_{n-1})~&=~
(p_n,{\tilde Q}{\tilde p}_{n-1})~=~ h_n~=~ {\tilde Q}_{n,-1} h_{n-1}
\cr
({\tilde P}^\dagger p_n,{\tilde p}_{n})~&=~
(p_n,{\tilde P}{\tilde p}_{n})~=~n h_n~=~{\tilde P}_{n,0} h_{n}
\cr}}

We obtain all the relations between the coefficients in \supdef, 
by expressing \commu\ and \opid\-\othp\ as
\eqn\expref{\encadremath{\eqalign{
[P,Q]~&=~Q-1\cr
[{\tilde Q}^\dagger,{\tilde P}^\dagger]~&=~{\tilde Q}^\dagger -1\cr
P+(1-Np)~&=~Ng\,{\tilde Q}^\dagger (Q-1) \cr
{\tilde P}^\dagger +(1-Nq)~&=~N g \, 
({\tilde Q}^\dagger -1)Q \cr}}}
and letting these operators act on $(p_n)$, namely
\eqn\letact{\eqalign{
Q_{n,k}-\delta_{k,0}~&=~\sum_{m\geq -1} Q_{n,m} P_{n-m,k-m}-P_{n,k-m}
Q_{n-k+m,m} \cr
{\tilde Q}_{n,k}-\delta_{k,0}~&=~\sum_{m\geq -1}
{\tilde P}_{n,k-m} {\tilde Q}_{n+k-m,m} -
{\tilde Q}_{n,m} {\tilde P}_{n+m,k-m} \cr}}
and
\eqn\lelet{\eqalign{
(1-Np)\delta_{k,0}+P_{n,k}+Ng {\tilde Q}_{n,-k}~&=~ 
Ng \sum_{m\geq -1} {\tilde Q}_{n-m,m-k} Q_{n,m}\cr
(1-Nq)\delta_{k,0}+{\tilde P}_{n,-k}+Ng Q_{n,k}~&=~ 
Ng \sum_{m\geq -1} {\tilde Q}_{n-m,m-k} Q_{n,m}\cr}}
This gives an infinite set of equations for the coefficients of
\supdef. 
A drastic simplification of \lelet\ follows from subtracting
the third from the fourth line of \expref, cancelling the
crossed term ${\tilde Q}^\dagger Q$. This results in
the operator identity
\eqn\main{A~=~P-Ng Q -Np~=~
({\tilde P}-Ng {\tilde Q}-Nq)^\dagger~=~ {\tilde A}^\dagger} 
where we have formed two convenient linear combinations
$A$ and $\tilde A$ of the $P$, $Q$ and $\tilde P$, $\tilde Q$ 
respectively.
When expressed on the basis $(p_n)$, \main\ yields
\eqn\mainpn{
P_{n,k}-Ng Q_{n,k}-Np \delta_{k,0}~=~
{\tilde P}_{n,-k}-Ng {\tilde Q}_{n,-k}-Nq \delta_{k,0}}
Taking $k>1$ (resp. $k<-1$) in this formula,
in which case the rhs (resp. the lhs) vanishes, this gives
\eqn\infim{ 
P_{n,k}~=~Ng \, Q_{n,k} \qquad {\tilde P}_{n,k}~=~Ng \, 
{\tilde Q}_{n,k}} 
for all $k>1$. 
Taking now $k=1$ and $k=-1$ in \mainpn\ leads to
\eqn\getko{\eqalign{
P_{n,1}- Ng\, Q_{n,1}~&=~- Ng\, {\tilde Q}_{n,-1} \cr
{\tilde P}_{n,1}- Ng\, {\tilde Q}_{n,1}~&=~- Ng\, Q_{n,-1} \cr}}
while $k=0$, \mainpn\ gives
\eqn\firnot{
Q_{n,0}-{\tilde Q}_{n,0}~=~{q-p \over g} }
But even after eliminating $P_{n,k}$,
${\tilde Q}_{n,k}$ and ${\tilde P}_{n,k}$, \letact\-\lelet\
remain an infinite set of quadratic equations for the $Q_{n,k}$.

In the next section, we shall derive a set of equations 
involving only the first values of the index $k$, at the cost
of taking derivatives wrt $g$.
We shall make use only of Eqs. \lead\ and \firnot\ ,
together with  the first line of \letact\ for $k=0$, which yields
\eqn\uko{ Q_{n,0} -1 ~=~ P_{n+1,1}-P_{n,1} }
This equation enables us to solve for $P_{n,1}$ in terms of the $Q_{i,0}$,
given that $P_{0,1}=0$ by definition. Introducing the quantity
\eqn\defvn{ v_n~=~ \sum_{k=0}^{n-1} Q_{k,0} }
we find that
\eqn\solpo{ P_{n,1}~=~v_n - n }
 
\subsec{The main recursion relation}

In this section, we take advantage of the simple dependence of
$Z$ on $g$ to derive differential equations for the first coefficients
$Q_{n,k}$, which, supplemented to the results of the previous section,
yield a recursion relation for $Z$.

Let us evaluate the following derivatives wrt $g$ (the notation
$\partial_g$ stands for $\partial/\partial g$)
\eqn\dereq{\eqalign{
\partial_g (p_n,{\tilde p}_{n-1})~&=~0~=~(\partial_g p_n,{\tilde p}_{n-1})
+(p_n,\partial_g {\tilde p}_{n-1})-N (Qp_n,{\tilde Q}p_{n-1})\cr
&=~(\partial_g p_n,{\tilde p}_{n-1})
-N ({\tilde Q}^\dagger Qp_n,{\tilde p}_{n-1})\cr
\partial_g (p_n,{\tilde p}_n)~&=~\partial_g h_n~=~
-N ({\tilde Q}^\dagger Qp_n,{\tilde p}_{n})\cr}}
where we have used the fact that $\partial_g p_n$ has degree $\leq n-1$, 
and the orthogonality of the $p$'s and $\tilde p$'s. To compute
$(\partial_g p_n,{\tilde p}_{n-1})$, let us write
$p_n(x)=x^n -\lambda_n x^{n-1}+ ...$ Using the definition of $Q$ \defqt,
we easily see that $\lambda_n=v_n$, given by \defvn. Indeed, we write
\eqn\wri{ \eqalign{
p_{n+1}(x)~&=~x^{n+1}- \lambda_{n+1} x^n +O(x^{n-1}) \cr
&=~Qp_n(x)-Q_{n,0}x^n +O(x^{n-1})\cr
&=~x^{n+1} -(\lambda_n+Q_{n,0})x^n+O(x^{n-1})\cr}}
and the result follows. Hence
\eqn\calpo{ 
(\partial_g p_n,{\tilde p}_{n-1})~=~-h_{n-1} \partial_g v_n }
Finally, let us express $N({\tilde Q}^\dagger Qp_n,{\tilde p}_m)$ by using
\opid, namely that $N{\tilde Q}^\dagger Q=N{\tilde Q}^\dagger 
+(P+1-Np)/g$, hence
\eqn\rexqp{\eqalign{
N({\tilde Q}^\dagger Qp_n,{\tilde p}_{n-1})~&=~
h_{n-1}\big(N{\tilde Q}_{n,-1} +
{1\over g}P_{n,1}\big)\cr
N({\tilde Q}^\dagger Qp_n,{\tilde p}_n)~&=~
h_n \big(N {\tilde Q}_{n,0}+{1\over g}
(P_{n,0}+1-Np)\big)\cr}}
and finally, using the explicit values \lead\ and \solpo, 
we get the differential equations
\eqn\evolu{\eqalign{
(\partial_g +{1 \over g}) v_n~&=~{n \over g} - N{h_n \over h_{n-1}}\cr
\partial_g \, {\rm Log}\, h_n~&=~{1\over g}(Np-n-1)
-N{\tilde Q}_{n,0}\cr
&=~{1\over g}(Nq-n-1)-N Q_{n,0}\cr}}
where we have used \firnot\ to get the last line.
Let us introduce
\eqn\frener{ F_n~=~ \sum_{k=0}^{n-1} {\rm Log} \, h_k }
so that 
\eqn\enerf{ {\rm Log}\, Z(p,q,g;N)~=~ F_N +{\rm Log}(e_N(g)N!) }
then after summation, the second equation of \evolu\ becomes
\eqn\becoe{ \partial_g F_n~=~{1\over 2g} n(2Nq-n-1) - Nv_n }
After elimination of $v_n$ between \evolu\ and \enerf, we are left with
\eqn\lefeq{ {1\over N^2}(\partial_g+{1\over g})\partial_gF_n~=~
{h_n\over h_{n-1}} -{n\over Ng}}                       
In terms of the variable $t=1/g$, and upon the definition
\eqn\aldef{\eqalign{ \alpha_n~&={h_n \over h_{n-1}}\qquad n=1,2,...\cr
\alpha_0~&=~ h_0 \cr}}
we finally have the system of equations
\eqn\fintr{\encadremath{\eqalign{
F_n~&=~\sum_{k=0}^{n-1} (n-k){\rm Log}\, \alpha_k\cr
\alpha_n~&=~n{t\over N}+{t^2 \over N^2}(t \partial_t)^2 F_n\cr}}}
Given the initial data $\alpha_0=h_0(p,q,1/t;N)$, 
this set of equation gives a recursion relation which determines 
$F_n$ for all $n$, and in particular $F_N$.

The normalization factor $e_N(g)$ in \enerf\ still has to be fixed.
It is obtained from the solution $(\alpha_n^{(0)},F_n^{(0)})$ of the system
\fintr\ for the case $p=q=0$. In that case, the initial data reads
\eqn\inicap{ \alpha_0^{(0)}~=~h_0(0,0,1/t;N)~=~
\int e^{-Nxy/t}dx dy~=~{t \over N}}
We then have 
\eqn\algen{ \alpha_n^{(0)}~=~ n{t \over N} }
for $n\geq 1$, and 
\eqn\frest{\eqalign{ F_n^{(0)}~&=~{\rm Log}\, {t \over N}+
\sum_{k=1}^{n-1} (n-k) {\rm Log}\left({kt\over N}\right)\cr 
&=~{\rm Log}\left( \left({t\over N}\right)^{n(n+1)/2}\prod_{j=0}^{n-1} j!
\right)\cr}}
Now the normalization $e_N(g)$ in \enerf\ ensures that $Z(0,0,1/t;N)=1$,
hence we have 
\eqn\final{ {\rm Log}\, Z(p,q,1/t;N)~=~F_N-F_N^{(0)} }
Upon comparison with \enerf, we deduce that
\eqn\valen{ e_N(1/t)~=~{1\over (t/N)^{N(N+1)/2} \prod_{j=0}^N j!} }

Introducing the renormalized quantities
\eqn\alreno{\eqalign{
a_n~&=~ {N \alpha_n \over n t}\qquad {\rm for}\ n=1,2,... \cr
a_0~&=~ {N\alpha_0 \over t} \cr}}
and the free energy
\eqn\fretru{ f_n~=~F_n-F_n^{(0)}} 
such that 
\eqn\finreZ{f_N~=~{\rm Log}\, Z(p,q,1/t;N)} 
we finally have the
recursive system, for $n\geq 1$
\eqn\recumain{\encadremath{\eqalign{
f_n~&=~\sum_{k=0}^{n-1} (n-k) {\rm Log}\, a_k \cr
a_n~&=~1+ {t\over N} (t \partial_t)^2 {f_n \over n} \cr}}}
with the initial data
\eqn\inidaa{\encadremath{\eqalign{ 
a_0~&=~{N\over t}\int dx dy (1-x)^{-Np} (1-y)^{-Nq}
e^{-Nxy/t} \cr
&=~ \sum_{k\geq 0} {1\over k!}{\Gamma(Np+k)\Gamma(Nq+k)\over \Gamma(Np)
\Gamma(Nq)}\left({t\over N}\right)^k \cr}}}
Note that if $-Np$ and/or $-Nq$ is a positive integer, the above sum truncates. 

At this stage, it is interesting to note that the quantity
$f_n$ which appears in the recursion relation \recumain\ has
the following simple interpretation. Indeed, for $n\ge 1$, we have 
\eqn\znf{f_n~=~{\rm Log}\, Z_n} 
where $Z_n=n!e_n(g)\prod_{i=0}^{n-1}h_i$
is the integral \pfunc\ taken over $n\times n$ matrices, but with the
{\it same} potential \pot , with $N$ as a prefactor.
The quantity $f_n$ is thus the corresponding free energy.
The normalization factor $e_n(g)$, which ensures that $Z_n=1$ for
$p=q=0$ is a straightforward extension of \valen 
\eqn\ovale{ e_n(g)~=~{(Ng)^{n(n+1)/2} \over \prod_{j=0}^n j!} }
In terms of fatgraphs, we now have a weight $n$ instead of $N$ per
oriented loop, leading to
\eqn\graxpn{ Z_n(p,q,1/t;N)~=~\sum_{{\rm fatgraphs}\ \Gamma}
{\left({n\over N}\right)^{F(\Gamma)}
t^{E(\Gamma)}\, N^{\chi(\Gamma)} \,  p^{V_p} 
q^{V_q} \over |{\rm Aut}(\Gamma)|} }
where $F(\Gamma)$ denotes the number of faces of the graph.
Therefore we may think of the new scaling variable $z=n/N$ as an activity
per {\it face} of the bi-colored graph.

\subsec{Solution of the main recursion relation}

To solve \recumain , we first eliminate $a_n$ by noting that 
\eqn\anfz{ a_n~=~{Z_{n+1} Z_{n-1} \over Z_n^2} }
for all $n\geq 0$ by setting $Z_{-1}=Z_0=1$. This is obtained 
by inverting the relations $Z_n=\prod_{k=0}^{n-1} a_k^{n-k}$.
In terms of $Z_n$, the recursion relation \recumain\ reads
\eqn\satiphi{\encadremath{ {Z_{n-1} Z_{n+1}\over Z_n^2}-1~=~
{t\over nN} (t\partial_t)^2 \, {\rm Log}\, Z_n }}
for all $n\geq 1$. The initial data are $Z_{-1}=Z_0=1$, and
$Z_1~=~a_0$ as in \inidaa.

Let us first solve \recumain\ in the case
\eqn\casepre{ a~=~Np~=~-1 }
and $b=Nq$ arbitrary. The initial data \inidaa\ reads 
\eqn\ezero{ a_0~=~1-b{t \over N} }
The solution of \satiphi\ takes the form
\eqn\valphi{ Z_n~=~1+\sum_{k=1}^n {n \choose k} b(b-1)...(b-k+1)
\left(-{t \over N}\right)^k }
for $n\geq 0$ and $Z_{-1}=1$, as easily checked 
by explicitly differentiating the rhs\foot{This result will also be 
proved directly in Sect.5.2 below.}.
In particular, the solution for $n=N$ reads
\eqn\solzet{ Z(-1/N,q,1/t;N)~=~1+\sum_{k=1}^N {N\choose k}
(Nq)(Nq-1)...(Nq-k+1) \left(-{t \over N}\right)^k }
A similar reasoning permits to solve the case
\eqn\socas{ a~=~Np~=~1}
and $b=Nq$ arbitrary, with now
\eqn\newphi{Z_n~=~1+\sum_{k=1}^\infty {n+k-1 \choose k} b(b+1)...(b+k-1)
\left({t\over N}\right)^k}
also satisfying \satiphi.
The corresponding solution for $n=N$ reads
\eqn\soloze{ Z(1/N,q,1/t;N)~=~1+\sum_{k=1}^\infty {N+k-1\choose k}
Nq(Nq+1)...(Nq+k-1) \left({t\over N}\right)^k }

More generally, we can use the recursion \satiphi\ to expand 
$f_n={\rm Log}\, Z_n(p,q,1/t;N)$ as a formal power series of $t$
\eqn\foropo{ f_n(p,q,1/t;N)~=~
\sum_{k\geq 1} \omega_{n,k}(a,b)\, \left({t \over N}\right)^k }
with $a=Np$, $b=Nq$.
The initial conditions $Z_0=1$ and $Z_1=a_0$ translate
into the conditions
\eqn\inifn{\encadremath{ \eqalign{ 
\omega_{0,k}(a,b)~&=~ 0\cr
\sum_{k\geq 1} \omega_{1,k} (t/N)^k~&=~{\rm Log}\, a_0 \cr} } }
with $a_0$ as in \inidaa.
The recursion relation \satiphi\ turns into a recursion relation for the
$\omega$'s
\eqn\recomeg{\encadremath{\eqalign{
\omega_{n+1,k}~&=~2\omega_{n,k}-\omega_{n-1,k}     \cr
&+\sum_{r\geq 1} {(-1)^{r-1} \over r\,  n^r} \sum_{k_1,...,k_r\geq 1\atop
\Sigma k_i= k-r} (k_1...k_r)^2 \omega_{n,k_1}...\omega_{n,k_r} \cr}}} 
Together with the initial conditions \inifn, this determines the 
coefficients $\omega_{n,k}(a,b)$ completely. It is easy to show
that these are polynomials of $a$, $b$, and $n$ for each value of $k$.
The first few of them read
\eqn\coefren{\eqalign{
\omega_{n,1}~&=~nab   \cr
\omega_{n,2}~&=~{nab \over 2}(n+a+b) \cr
\omega_{n,3}~&=~{n a b\over 3}(n^2+3(a+b)n+a^2+3 a b+b^2+1)\cr
\omega_{n,4}~&=~
{n a b\over 4}(n^3+6(a+b)n^2+(6a^2+17 ab +6 b^2+5)n\cr
&+(a+b)(a^2+5 ab+b^2+5))\cr
\omega_{n,5}~&=~
{n ab\over 5}(n^4+10(a+b)n^3+5(4a^2+11 ab+4 b^2+3)n^2\cr
&+5 (a+b)(2 a^2+9 ab+2 b^2+8)n\cr
&+a^4+10 a^3b+20 a^2b^2+10 ab^3+b^4+15a^2+40 ab+15 b^2+8)\cr}}

\fig{The construction of the tri-colored graph $\tilde \Gamma$ (b)
corresponding to the bi-colored graph $\Gamma$ (a). 
Each face is subdivided into triangles, by adding a new central vertex,
colored in grey. The face-weight
$z$ is re-affected to each such new grey vertex.}{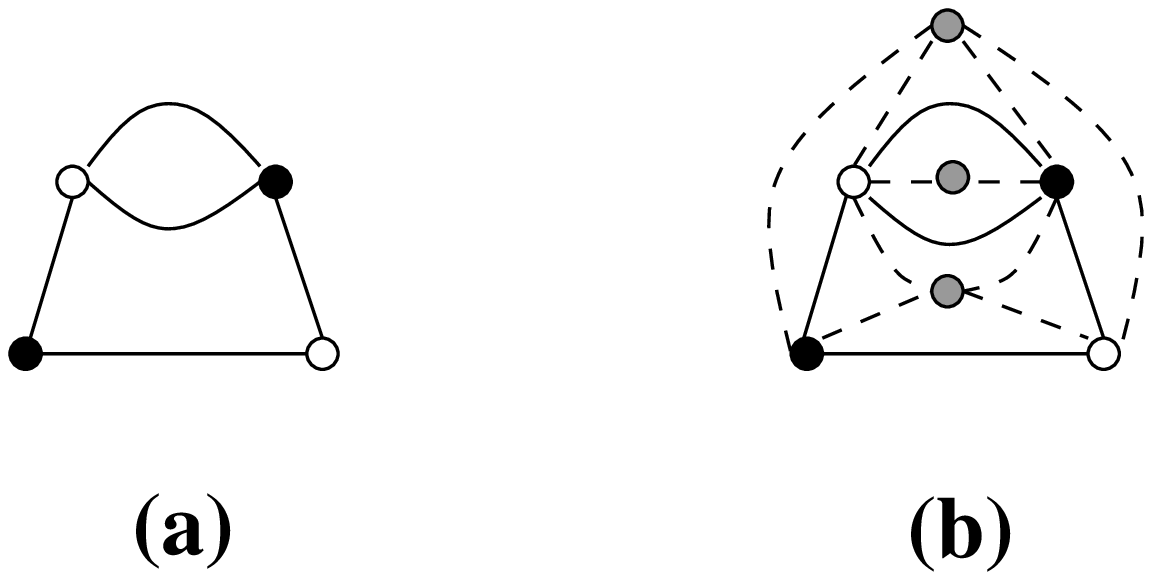}{8.cm}
\figlabel\sypqz

Remarkably, the result \coefren\ reveals that each
coefficient $\omega_{n,k}(a,b)$ is {\it totally symmetric} in 
the three variables $(a,b,n)$, or equivalently the three variables
$(p,q,z=n/N)$. Let us prove this property directly on
$f_n(p,q,1/t;N)$ \foropo\ by considering its fatgraph expansion \graxpn\ 
(restricted to connected graphs). For each fatgraph $\Gamma$ with black and white 
vertices in the expansion \graxpn, let us construct one
grey vertex in the middle of each face 
(see Fig.\sypqz), and connect it to each  
(black or white) vertex around the face. As black and white vertices
alternate around each face, this creates an even number of edges.
Each grey vertex receives a contribution $z$, 
whereas white vertices are weighed by $p$ and black ones by $q$.
We end up with a fatgraph $\tilde \Gamma$ whose vertices are tri-colored
(in white, black and grey),
and whose faces are all triangles with a vertex of each color. The
automorphism group remains unchanged ($|$Aut$({\tilde \Gamma})|=|$Aut$(\Gamma)|$), 
so as the Euler characteristic ($\chi({\tilde \Gamma})=\chi(\Gamma)$), 
hence we get the expansion
\eqn\expgrt{ f_n(p,q,1/t;N)~=~\sum_{{\rm conn.}\ {\rm tricol.}\atop
{\rm fatgraphs}\ {\tilde \Gamma} } t^{F\over 2}\, 
{p^{V_w}\, q^{V_b} \, z^{V_g} \,N^{\chi} \over |{\rm Aut}({\tilde \Gamma})|}}
where $F$ denotes the total number of (triangular) faces of $\tilde \Gamma$
($F=2E(\Gamma)$), 
and $V_w$, $V_b$, $V_g$ respectively its numbers of white, 
black and grey vertices.
The form \expgrt\ is explicitly symmetric in the three variables
$(p,q,z)$, that is in the three variables $(a,b,n)$.

The $a \leftrightarrow b$ symmetry was clear from the integral \pfunc.
Another direct proof of the symmetry $n \leftrightarrow a$ will be
given in Sect.4.2 below.
As an additional check, we may compare $Z_1(a=n,b)=a_0(a=n,b,t)$ as given
by \inidaa, to $Z_n(1,b)$ found in \newphi: the two sums are
indeed identical.

\subsec{Large $N$ limit}

In the large $N$ limit, the free energy Log$\, Z$ is dominated
by the planar (genus zero) fatgraphs, hence a leading contribution
\eqn\leaerg{{\rm Log}\, Z(p,q,1/t;N)~\sim~ N^2 F_0(p,q,t)}
when $N\to \infty$.
Introducing the rescaled
variable $z=n/N$, we may assume that $a_n$ tend to a smooth
function $a(z;p,q,t)$ of $z\in [0,1]$ in the large $N$ limit. This is
in agreement with the assumption that $f_n/N^2$ also tends
to a smooth function $f(z;p,q,t)$. Indeed, \recumain\ then implies that
\eqn\larN{ a(z;p,q,t)~=~1+ {t\over z} (t\partial_t)^2 f(z;p,q,t) }
According to \graxpn, the continuous quantity $z$ is nothing but  
a fugacity associated with the faces of the (planar) fatgraphs.

The recursion relation \satiphi\ reads, for $\epsilon=1/N$
and $f(z)\equiv f(z;p,q,t)$
\eqn\repsz{ e^{N^2(f(z+\epsilon)+f(z-\epsilon)-2 f(z))+O(1)}-1~=~
{t\over nN} (t\partial_t)^2 (N^2 f(z)+O(1))}
hence in the limit $N\to \infty$
\eqn\diffr{\encadremath{ \partial_z^2 f~=~
{\rm Log}(1+{t\over z} (t\partial_t)^2 f)}}
The partial differential equation \diffr\ 
however does not determine $f(z;p,q,t)$ completely.
We need to supplement it by some initial condition. This is easily done
by use of the formal small $t$ series expansion of $f$. From the
large $N$ limit of \foropo\-\coefren\ (with $a=Np$, $b=Nq$), 
we indeed find that
\eqn\fitha{\eqalign{ f_n~&=~nab{t\over N}+ O(t^2)\cr
&=~N^2\, {pqz}\, t +O(t^2) \cr}}
hence we get the initial condition
\eqn\inicon{ f(z;p,q,t)~=~ {pqz}\, t +O(t^2) }
This can be used to feed the differential equation \diffr. 
Indeed, writing
\eqn\fnwr{ f(z;p,q,t)~=~\sum_{k\geq 1} \omega_k(z;p,q) t^k }
with $\omega_1=pqz$, 
the equation \diffr\ turns into a recursion relation
\eqn\recomge{ \partial_z^2 \omega_k~=~\sum_{r\geq 1} 
{(-1)^{r-1}\over r\, z^r} 
\sum_{k_1,...,k_r\geq 1
\atop \Sigma k_i=k-r} k_1^2 ... k_r^2 \, \omega_{k_1} ...\omega_{k_r}}
for $k\geq 2$. This is nothing but the continuum limit of the recursion 
relation \recomeg. Eq.\recomge\ however involves two integrations wrt 
$z$ at each step, hence the introduction of many integration constants.
To fix them, we need to know the values of $f(0;p,q,t)$ and
$\partial_z f(0;p,q,t)$. The latter are readily obtained by
analyzing the large $N$ limit of the initial condition
\inidaa, namely we write
\eqn\larn{\eqalign{ Z_1~&=~a_0~=~\sum_{k\geq 0}
{1\over k!}{\Gamma(Np+k)\Gamma(Nq+k)\over \Gamma(Np)
\Gamma(Nq)}\left({t\over N}\right)^k \cr
&\sim N\int_0^\infty d\alpha\, e^{N\, S(\alpha;p,q,t)}\cr}}
where we have replaced the summation over $k=\alpha N$ by
an integral over the positive real numbers $\alpha$, and
used the Stirling formula to derive the effective
action
\eqn\efac{\eqalign{S(\alpha;p,q,t)~&=~
(p+\alpha)({\rm Log}(p+\alpha)-1)
+(q+\alpha)({\rm Log}(q+\alpha)-1)+\alpha\, {\rm Log}\, t \cr
&-\alpha({\rm Log}\, \alpha -1)
-p({\rm Log}\, p-1)-q({\rm Log}\, q-1) \cr}}
The integral is dominated by the saddle-point $dS/d\alpha(\alpha^*)=0$,
namely
\eqn\sappo{ t(\alpha^*+p)(\alpha^*+q)~=~\alpha^* }
We only retain the solution leading to a maximum of $S$, namely
\eqn\sapo{ \alpha^*~=~ {1-t(p+q)\over 2t }
-{1\over 2t}\sqrt{(1-t(p+q))^2-4pqt^2} }
Finally, the free energy $f_1~=~{\rm Log}\, Z_1$ behaves for large
$N$ as
\eqn\frenone{\eqalign{ f_1(p,q,t)~&\sim~ N\, S(\alpha^*) \cr
&=~N\, (
p\, {\rm Log}(1+{\alpha^*\over p})+
q\, {\rm Log}(1+{\alpha^*\over q})-\alpha^*) \cr
& \equiv N\, \varphi_1(p,q,t)\cr }}
Let us now interpret the result \frenone. As $n=1$, we have $z=1/N$.
The fact that $f_1$ is of order $N=N^2 z$ shows that
\eqn\interp{ f(z;p,q,t)~=~ z \varphi_1(p,q,t)+ O(z^2) }
for small $z$, hence that
\eqn\inidataf{\eqalign{
f(0;p,q,t)~&=~0\cr
\partial_z f(0;p,q,t)~&=~ \varphi_1(p,q,t)~
=~\sum_{k\geq 1} \mu_k(p,q)\, t^k\cr}}
with $\varphi_1$ as in \sapo\-\frenone. 
The coefficients $\mu_k(p,q)$
are computed in Appendix A, and read
\eqn\valmuk{ \mu_k(p,q)~=~{1\over k^2}\sum_{j=1}^k {k \choose j}
{k \choose j-1} \, p^j \, q^{k+1-j} }
The quantity $\varphi_1(p,q,t)$ can be viewed as the free energy
associated with bi-colored graphs with exactly one vertex, i.e.
bi-colored trees. As clear from \sapo , it develops a singularity
at a critical value for $t$ equal to
\eqn\tstarzzero{t_*(p,q,0)={1\over \left(\sqrt{p}+\sqrt{q}\right)^2}}
For $t\to t_*(p,q,0)$, we find (see Appendix A)
\eqn\tapproa{\varphi_1(p,q,t)\sim (t_*-t)^{3/2}}
As we shall see however, this behavior for the free energy,
valid only in the limit $z\to 0$, is not generic.

With the initial data \inidataf\-\valmuk , we may
now write the differential equation \recomge\ in integrated form
with $\omega_k(z)\equiv\omega_k(z;p,q)$
\eqn\newrecom{
\omega_k(z)~=~z \, \mu_k(p,q) +\int_0^z dx \int_0^x dy
\sum_{r\geq 1} {(-1)^{r-1}\over r\, y^r}
\sum_{k_1,...,k_r\geq 1
\atop \Sigma k_i=k-r} k_1^2 ... k_r^2 \,
\omega_{k_1}(y) ...\omega_{k_r}(y)}
This is now a true recursion relation on $k$, which determines all
$\omega$'s.
With the
initial condition $\omega_0(z;p,q)=0$, we find
\eqn\finome{\eqalign{
\omega_1~&=~pqz \cr
\omega_2~&=~{pqz\over 2}(z+p+q)     \cr
\omega_3~&=~{pqz\over 3}(z^2+3(p+q)z+p^2+3 pq+q^2)  \cr
\omega_4~&=~{pqz\over 4}(z^3+6(p+q)z^2+(6p^2+17pq+6q^2)z+
(p+q)(p^2+5 p q +q^2))\cr
\omega_5~&=~{pqz\over 5}(z^4+10(p+q)z^3+(20p^2+55pq+20q^2)z^2\cr
&+5(p+q)(2p^2+9pq+2q^2)z+p^4+10 p^3q+20p^2q^2+10pq^3+q^4)\cr
\omega_6~&=~{pqz\over 6}(z^5+15 (p+q) z^4+5 (10 p^2+27 p q+10 q^2)
z^3\cr
&+2 (p+q)(25 p^2+106 p q+25 q^2) z^2+
(15 p^4+135 p^3 q+262 p^2 q^2+135 p q^3+15 q^4) z \cr
&+(p+q)(p^4+14 p^3 q+36 p^2 q^2+14 p q^3+q^4)) \cr}}
For $z=1$, this gives the first few terms in the expansion
of the planar free energy
\leaerg\ as a formal power series of $t$. In this expansion,
the coefficient of $t^k p^r q^s$ corresponds to the planar
connected (dual) black and white fatgraphs
with $r$ white faces, $s$ black faces, and $k$ edges separating them,
weighed by the inverse of the order of their symmetry group.
For $z$ arbitrary, the coefficient of $z^m$ singles
out those black and white graphs with exactly $m$ vertices. 
Since we are dealing with planar graphs, one has $m+r+s=k+2$.
This explains why $\omega_k(z;p,q)$ is a {\it homogeneous}
polynomial of $p$, $q$, $z$, of degree $k+2$.
Note also that \finome\ agrees with the large
$N$ limit of \coefren, i.e.
\eqn\limite{ \lim_{N\to \infty} {1 \over N^2} \omega_{zN,k}(Np,Nq)~=~
\omega_k(z;p,q)}
Finally, the result \finome\ is explicitly symmetric in the three variables
$(p,q,z)$, in agreement with the expected symmetry
\eqn\sympqz{ f(z;p,q,t)~=~f(z;q,p,t)~=~f(p;z,q,t) }

\newsec{Determinant form and discrete Hirota equation}

\subsec{The partition function as a determinant}

Throughout this section, we use the shorthand
notation $Z_n(a,b)$ for the
partition function $Z_n(p,q,g;N)$ \pfunc\ for an integral over 
$n\times n$ matrices, and with $a=Np$, $b=Nq$.

We start from the reduced integral \redint\-\measu, extending over
the $2n$ eigenvalues $m_i$ and $r_i$
\eqn\inforp{ Z_n(a,b)~=~e_n(g)\int \prod_{i=1}^n dm_i dr_i
(1-m_i)^{-a} (1-r_i)^{-b} \Delta(m) \Delta(r) e^{-Ng\sum_{i=1}^n m_i r_i}}
{}From the definition of the Vandermonde determinant,
we have $\Delta(1-m)=(-1)^{n(n-1)/2} \Delta(m)$, hence we may replace 
$m\to 1-m$ and $r\to 1-r$ in both Vandermonde determinants without
altering \inforp. Moreover,
\eqn\detergent{\eqalign{ 
\prod_{i=1}^n (1-m_i)^{-a} \Delta(1-m)~&=~\det\, \bigg[
(1-m_i)^{j-a-1} \bigg]_{1\leq i,j \leq n} \cr
&=~\sum_{\sigma \in S_n} {\rm sgn}(\sigma) \prod_{i=1}^n 
(1-m_i)^{\sigma(i)-a-1}\cr}}
and we have an analogous formula for $r$ (with $b$ instead of $a$).
Substituting these expansions into \inforp, we get 
\eqn\getzn{\eqalign{ Z_n(a,b)~&=~e_n(g)\sum_{\sigma,\tau\in S_n}
{\rm sgn}(\sigma\tau)\cr
&\times\prod_{i=1}^n \int dm_i dr_i
(1-m_i)^{\sigma(i)-a-1} (1-r_i)^{\tau(i)-b-1} e^{-Ngm_i r_i}\cr
&=~n!e_n(g)\sum_{\nu\in S_n}{\rm sgn}(\nu)\prod_{i=1}^n
\int dx dy (1-x)^{i-a-1} (1-y)^{\nu(i)-b-1}e^{-Ngxy} \cr}}
where we have set $\nu=\tau \sigma^{-1}$, with the same signature 
as $\sigma\tau$, and explicitly factored out the sum over $\sigma$.
Moreover, the dummy integration variables have been rebaptized 
$x$ and $y$. The partition function takes therefore the form
\eqn\deterfor{ Z_n(a,b)~=~n!e_n(g)\, \phi_n(a,b) }
where $\phi_n(a,b)$ is the $n\times n$ determinant
\eqn\phidefi{\encadremath{
\phi_n(a,b)~=~\det\bigg[ \int dx dy (1-x)^{i-a-1}
(1-y)^{j-b-1} e^{-Ngxy} \bigg]_{1\leq i,j \leq n} }}

\subsec{Discrete Hirota equation}

The technique used in this section to derive a discrete Hirota
equation for $\phi_n$ is borrowed from \KAZ. The argument is based
on a general quadratic relation satisfied by the minors of the determinant
$D$ of any
$(n+1)\times (n+1)$ matrix. If we denote by $D_{i,j}$ the $n\times n$ minor
of $D$ obtained by erasing the $i$-th row and $j$-th column, and
$D_{i_1,i_2;j_1,j_2}$ the $(n-1)\times (n-1)$ minor 
obtained by removing the rows $i_1,i_2$
and columns $j_1,j_2$, we have the relation
\eqn\qadra{ D\, D_{1,n+1;1,n+1} ~=~ D_{n+1,n+1}\, D_{1,1} - D_{1,n+1}\, D_{n+1,1}}
This is proved for instance by collecting the monomials in the expansions 
of the various minors. 
This may also be viewed as a particular case of the Pl\" ucker relations
\WIG.
When applied to $\phi_{n+1}(a+1,b+1)$, \qadra\ immediately
translates into the quadratic relation
\eqn\hiro{ \phi_{n+1}(a+1,b+1)\phi_{n-1}(a,b)~=~\phi_n(a+1,b+1)\phi_n(a,b)-
\phi_n(a,b+1)\phi_n(a+1,b) }
where we have absorbed the various restrictions on the matrix indices $i,j$
into corresponding shifts of $a$ and $b$, while the index of $\phi$ 
refers to the size of the minor.
The equation \hiro, is known as a discrete Hirota equation, playing
a central role in integrable systems (see \WIG\ for the study of
analogous equations).

We may recast \hiro\ in terms of $Z_n$, using \phidefi\ and
the value \ovale\ of $e_n(g)$ satisfying 
\eqn\sotha{ {(n+1)!e_{n+1}(g) (n-1)!e_{n-1}(g) \over (n!e_n(g))^2}~
=~ {1\over Ng} {n! \over (n-1)!}~=~{n\over Ng} }
This leads to 
\eqn\hiroz{\encadremath{\eqalign{ 
n {t\over N}\, Z_{n+1}(a+1,b+1)\, &Z_{n-1}(a,b)\cr
&=~ Z_n(a+1,b+1)\, Z_n(a,b)-Z_n(a,b+1)\, Z_n(a+1,b)\cr}}}
The equation \hiroz\ gives a completely 
different information on $Z_n(a,b)$ or $f_n(a,b)$
than the recursion relation \satiphi. 
Indeed, the latter determines
the dependence of $Z_n$ on $a$, $b$ from the initial data $Z_1$ and
through a recursive process, the former instead gives a three-dimensional
recursion relation on the parameters $a$, $b$ and $n$.
In terms of the free energy 
\eqn\freab{ f_n(a,b)~=~ {\rm Log}\, Z_n(a,b)}
and using the difference operator 
\eqn\difopera{ \delta_x f(x)~=~ f(x+1)-f(x) }
we can rewrite \hiroz\ as
\eqn\rewiro{\encadremath{ 
\delta_a\delta_b f_n(a,b)~=~-{\rm Log}\big(1-n {t\over  N}
e^{\delta_n f_n(a+1,b+1) -\delta_n f_{n-1}(a,b)}
\big)}}
Substituting for $f_n(a,b)$ the small $t$ expansion \foropo, this
gives yet another recursion relation for the 
coefficients $\omega_{n,k}(a,b)$, with the form
\eqn\recomeg{ \delta_a \delta_b\omega_{n,k+1} ~=~
\varphi(\omega_{j,l}| j\in \{n+1,n,n-1\}\, ;\,  l\leq k)}
This is a recursion over $k$, with the initial conditions
$\omega_{n,0}=0$ and $\omega_{n,1}=nab$. However, it involves
a discrete integration wrt $a$ and $b$, namely we have to solve
an equation of the form $\delta_a\delta_b l=h$ at each step.
As $h$ is a polynomial of $a$ and $b$, this is 
readily done by noticing that the polynomial $\omega_{n,k}(a,b)$
is always a factor of $ab$. Indeed, from the original 
interpretation  of $f_{n=N}$ as a sum over connected fatgraphs
as the logarithm of \graxp,
each connected fatgraph must have at least one $M$ and one $R$-vertex,
resulting in a factor $ab$. For $n\neq N$, an analogous expansion holds,
and the same conclusion applies.
In this way, we recover the results \coefren. Note that the step 
of discrete integration is simpler here than in \recomeg, where
the two discrete integrations wrt $n$ actually involved the determination
of $\omega_{n,k}$ for all $n=0,1,2,...$. Here this step can be done 
in general without having to compute explicitly for all (integer) values
$a,b=0,1,2,...$ It is much more economical when implemented on a computer.

\subsec{Large $N$ limit}

In the planar $N\to \infty$ limit, we still set $z=n/N$. 
We also assume the leading behavior $f_n\sim N^2 f(z;p,q,t)$,
so that the large $N$ free energy \leaerg\ reads $F_0(p,q,t)=f(1;p,q,t)$.
The
above difference operators \difopera\
become differential operators, namely
\eqn\diffN{ \delta_n \to {1 \over N} \partial_z \qquad
\delta_a \to {1\over N}\partial_p \qquad \delta_b
\to {1\over N} \partial_q}
and the Hirota equation \rewiro\ becomes a partial differential equation
for the function $f\equiv f(z;p,q,t)$
\eqn\pardiro{\encadremath{\partial_p\partial_q f~=~
-{\rm Log}\big(1-tz e^{\partial_z(\partial_z +\partial_p+\partial_q)f}\big) }}
Note that $f$ is not determined entirely by this equation, in particular
we may add to $f$ any function of $t$. However, as mentioned earlier,
$f$ has $pq$ in factor, which implies that $\partial_p f(z;0,0,t)=
\partial_q f(z;0,0,t)=0$. 
We therefore solve an equation of the form 
$\partial_p\partial_q f(p,q)=g(p,q)$ as
$f=\int_0^pdx \int_0^qdy\, g(x,y)$ (we will
actually implement this on the
coefficients $\omega_k(z;p,q)$ of the expansion \foropo, 
which are polynomials of $p$, $q$ for all $k$).
Substituting the small $t$ expansion \fnwr\
into \pardiro, and integrating wrt $p$ and $q$, we finally obtain 
\eqn\obtom{\sum_{n\geq 0}\omega_n(z;p,q)\, t^n~=~-\int_0^p dx\int_0^q dy\,
{\rm Log}\big( 1 -tze^{\sum_{k\geq 0} 
\partial_z(\partial_z+\partial_x+\partial_y)\omega_k(z;x,y) t^k}\big) }
which is a recursion relation for $\omega_n$. With the
initial condition $\omega_0(z;p,q)=0$, we recover the result 
\finome.
Note that both \diffr\ and \pardiro\ are partial differential equations
determining $f$ up to initial conditions, they must therefore
be compatible, a highly non-trivial fact considering their very
different forms.

Remarkably, combining both equations \diffr\ and \pardiro\ 
allows us to solve the case $p=q=z$ explicitly, 
with the following result for the reduced free energy\foot{
This result coincides with the generating function of planar
rooted bicubic maps of $2n$ vertices \TUT.} 
\eqn\redufr{\encadremath{\eqalign{ F(z,t)~&=~f(z;z,z,t)\cr
&=~3\, z^2\, \sum_{k=1}^\infty (2tz)^k \, {(2k-1)! \over k! (k+2)!} 
\cr}}}
as we will prove now. 
The $(p,q,z)$ symmetry \sympqz\ of $f$ implies that
\eqn\impsym{
\partial_z(\partial_z+\partial_p+\partial_q)f(z;z,z,t)~=~
(\partial_z^2+2\partial_p\partial_q)f(z;z,z,t)~=~
{1 \over 3} \partial_z^2 F(z,t) }
Now using the equations \diffr\ and \pardiro\ in the form
\eqn\forpari{\eqalign{ \partial_z^2f~&=~{\rm Log}\big(
1+{t\over z}(t\partial_t)^2f\big)\cr
\partial_p\partial_q f ~&=~-{\rm Log}\big(1-zt 
e^{\partial_z(\partial_z+\partial_p+\partial_q)f}\big)\cr}}
and adding twice the second line to the first one, we get, using
\impsym , the equation satisfied by $F(z,t)$
\eqn\weget{\eqalign{ 
{1\over 3}\partial_z^2 F(z,t)~&=~{\rm Log}\big(1+{t\over z}
(t\partial_t)^2F(z,t)\big) \cr
&-2 \, {\rm Log}\big(1-zt e^{{1\over 3}\partial_z^2F(z,t)}\big)\cr}}
To check that \redufr\ is a solution of this equation,
we can relate the expression
\eqn\exfdp{ {1 \over 3} \partial_z^2 F(z,t)~=~\sum_{k=1}^\infty
(2tz)^k \, {(2k-1)! \over (k!)^2} }
to the generating function of the celebrated
Catalan numbers
\eqn\catagenera{ C(x)~=~
{1 - \sqrt{1-4x} \over 2x}~=~\sum_{k\geq 0} c_k\, x^k}
with
\eqn\catalan{ c_k~=~ {(2k)!\over k! (k+1)!} }
Setting $x=2zt$, we may reexpress the result \exfdp\ as
\eqn\rexfdt{\eqalign{ 
{1\over 3} \partial_z^2 F(z,t)~&=~
\int_0^x {dy\over 2y} \big({d \over dy} (yC(y))-1 \big) \cr
&=~\int_0^x {dy \over 2y}\bigg({1 \over \sqrt{1-4y}}-1 \bigg)\cr 
&=~{1\over 2}{\rm Log}\left({1-\sqrt{1-4x} \over 1+\sqrt{1-4x}}\right)
-{1\over 2}{\rm Log}\, x\cr
&=~{\rm Log}\, C(x) \cr}}
Exponentiating both sides of \weget\ and using \rexfdt, our result will hold
provided 
\eqn\provid{ C(x) \big(1-{x\over 2}C(x)\big)^2~=~ 1+{t\over z}
(t\partial_t)^2 F(z,t) }
with $x=2zt$ as above. But thanks to the  quadratic relation
$xC(x)^2=C(x)-1$ satisfied by \catagenera, we may rewrite the lhs
of \provid\ as
\eqn\viprod{\eqalign{ 
1+{x^2\over 4} C(x)^3~&=~1+{x\over 4}C(x)(C(x)-1)\cr
&=~1+{1\over 4}((1-x)C(x)-1) \cr
&=~1+\sum_{k=1}^\infty {x^k\over 4}(c_k-c_{k-1})\cr
&=~1+{3\over 4}\sum_{k=0}^\infty (2tz)^{k+1} 
{(2k)!\over (k-1)!(k+2)!}\cr
&=~1+3tz(t\partial_t)^2\sum_{k=0}^\infty (2tz)^k 
{(2k-1)!\over k!(k+2)!}  \cr}} 
This proves \provid\ for $F(z,t)$ as in \redufr . Since
\redufr\ also satisfies the correct initial conditions
$F(0,t)=f(0;0,0,t)=0$ and $\partial_zF(0,t)=3\partial_zf(0;0,0,t)=0$,
it is therefore identified as the desired solution for $p=q=z$.
Note that, from the convergence radius of the series \redufr ,
this solution has a critical point at
\eqn\tstarzzz{t_*(z,z,z)={1\over 8z}}

Actually, the generating function at $p=q=z=1$
for bi-colored fatgraphs with a {\it marked} edge reads 
(see Appendix A for a similar study)
\eqn\markedge{ (t\partial_t)F(1,t)~=~ \sum_{k=1}^\infty t^k \,
\nu_k }
where 
\eqn\nuk{ \nu_k~=~{3\over 2}\,  2^k \, {c_k \over k+2} }
Remarkably, the numbers $\nu_k$ have already emerged in the context 
of arches and meanders \DGGB, within the framework of the 
Temperley-Lieb algebra. A more detailed study would indeed lead to 
the identification between certain pairs of arch configurations of
order $k$ (with $k$ arches) and bi-colored diagrams with $k$ edges, 
thus yielding a different interpretation of the result \nuk.

\newsec{Direct expansion}

\subsec{The partition function as a sum over integers}

We start from the formula \phidefi\ for the 
$n\times n$ determinant $\phi_n(a,b)$. For generic
(non-integer) values of $a$ and $b$, we may expand
\eqn\expanab{ (1-x)^{i-a-1}~=~ \sum_{k\geq 0} 
{\Gamma(k+a-i+1)\over \Gamma(1+a-i)} {x^k \over k!} }
and similarly for $(1-y)^{j-b-1}$. Using the formal integral
\eqn\forintegr{ \int dx dy \, x^\alpha y^\beta e^{-Nxy/t}~=~
\alpha! \, \delta_{\alpha,\beta}\, (t/N)^{\alpha+1} }
for any integer $\alpha$, we get
\eqn\redet{\eqalign{
\phi_n&(a,b)~=~\det\bigg[\sum_{k\geq 0}{1\over k!} 
{\Gamma(k+a-i+1)\over \Gamma(1+a-i)}
{\Gamma(k+b-j+1)\over \Gamma(1+b-j)} \big( {t\over N}\big)^{k+1}
\bigg]_{1\leq i,j\leq n} \cr
&=~\sum_{k_1,...,k_n\geq 0} \prod_{i=1}^n 
{1\over k_i!}{\Gamma(k_i+a-i+1)\over \Gamma(1+a-i)}
{(t/N)^{k_i+1}\over \Gamma(1+b-i)} \det\big[\Gamma(k_i+b-j+1)
\big]_{1\leq i,j \leq n}\cr}}
where we have used the multilinearity of the determinant to 
extract line by line the summations over $k$'s.
Factoring $\Gamma(k_i+b-n+1)$ out of each line (number $i$)
of the remaining determinant, we are left with a determinant 
of the form
\eqn\deterleftz{ \det \big[(k_i+b-j)(k_i+b-j-1)...(k_i+b-n+1)\big]~=~
\det \big[ q_{n-j}(k_i) \big] }
where the polynomials $q_m(x)=x^m+$lower degree are monic.
As in \monic, we may identify \deterleftz\ as the Vandermonde determinant
$\Delta(k)$ of the diagonal matrix $k=$diag$(k_1,k_2,...,k_n)$.
This yields
\eqn\yiedet{ 
\phi_n(a,b)~=~
\sum_{k_1,...,k_n\geq 0} \prod_{i=1}^n 
(t/N)^{k_i+1}{1\over k_i!}
{\Gamma(k_i+a-i+1)\over \Gamma(1+a-i)}
{\Gamma(k_i+b-n+1)\over \Gamma(1+b-i)} \Delta(k) }
Using the antisymmetry of $\Delta(k)$, we may symmetrize the above
expression into
\eqn\simdet{\eqalign{
\phi_n(a,b)~&=~
{1\over n!}\sum_{k_1,...,k_n\geq 0} \prod_{i=1}^n {1\over k_i!}
{(t/N)^{k_i+1}\over \Gamma(1+a-i)}
{\Gamma(k_i+b-n+1)\over \Gamma(1+b-i)} \cr
&\times \Delta(k)\, \det[
\Gamma(k_i+a-j+1)]_{1\leq i,j \leq n}\cr} }
Factoring again $\Gamma(k_i+a-n+1)$ for each line (number $i$)
of the last determinant, 
and repeating the above trick, we finally get
\eqn\finphidet{
\phi_n(a,b)~=~{1\over n!}
\sum_{k_1,...,k_n\geq 0}\Delta(k)^2 \prod_{i=1}^n 
(t/N)^{k_i+1}{1\over k_i!}{\Gamma(k_i+a-n+1)\over \Gamma(1+a-i)}
{\Gamma(k_i+b-n+1)\over \Gamma(1+b-i)}}
or, using \deterfor, 
\eqn\zndetf{
Z_n(a,b)~=~
\sum_{k_1,...,k_n\geq 0} \Delta(k)^2 \prod_{i=1}^n 
{(t/N)^{k_i+1-i}\over i!}{1\over k_i!}{\Gamma(k_i+a-n+1)\over \Gamma(1+a-i)}
{\Gamma(k_i+b-n+1)\over \Gamma(1+b-i)} }
The expression summed over is symmetric in the $k_i$'s, hence 
we may further reduce the sum to only strictly increasing
sequences of $k_i$'s, namely
\eqn\order{\encadremath{\eqalign{
&Z_n(a,b)~=~\cr
&\sum_{0\leq k_1<k_2...<k_n} \Delta(k)^2 \prod_{i=1}^n 
{(t/N)^{k_i-(i-1)}\over (i-1)!}{1\over k_i!}
{\Gamma(k_i+a-n+1)\over \Gamma(1+a-i)}
{\Gamma(k_i+b-n+1)\over \Gamma(1+b-i)} \cr}}}
This expression leads to a straightforward expansion of $Z_n$
in powers of $t$.

The leading $t^0$ term corresponds to $k_i=i-1$, which we refer to as
the {\it fundamental} (or {\it groundstate}) configuration of the
$k_i$'s. It has the contribution $1$.  The term of order one in $t$
is obtained by creating the excited state where $k_i$ are unchanged
for $i=1,...,n-1$ and $k_n=n-1 \to k_n=n$. Its contribution is easily
computed as $n a b t/N$. We may think of the power of $t$ as the 
energy of the excited state ($1$ here).
In general the term of order $t^k$ will be obtained by a number 
of excitations of the form $k_i=i-1+\delta k_i$, which respect the
strict ordering of the $k's$ (i.e., $\delta k_i\le\delta k_{i+1}$
for all $i$), and with a total energy 
$k=\sum_{i}\delta k_i$. This makes the expansion of $Z_n(a,b)$
quite explicit.

\subsec{Explicit Expansions}

As an illustration of the use of \order, let us re-work the example
$a=-1$,
$b$ arbitrary, of \solzet, by using the expansion \order.
The terms $\Gamma(1+a-i)=\Gamma(-i)$, $i=1,2,...,n$, in the denominator
may cause the contribution to vanish, {\it unless}
they are counterbalanced by terms of the form $\Gamma(-j)$,
$j \geq 0$ coming from the numerator. Hence, the only non-zero
contributions to the sum \order\ arise when
$k_i+a-n+1=k_i-n\leq 0$, for all $i$. The $k$'s being strictly ordered,
they are either in the fundamental configuration
\eqn\fundacon{K_0~=~ (k_1,k_2,k_3,...,k_n)~=~(0,1,2,3,...,n-1) }
or in one of the following $n-1$ excited states
\eqn\excits{ K_j~=~ (0,1,...,j-2,j-1,j+1,j+2,...,n)}
for $j=1,2,...,n-1$.
Assembling the contributions from the configurations
$K_0,K_1,...,K_{n-1}$, we get
\eqn\resuy{ Z_n(-1,b)~=~1+\sum_{j=1}^{n-1} \left({t \over N}\right)^{n-j}
{n \choose j} {\Gamma(1+b) \over \Gamma(1+b-n+j)} }
which is equivalent to \valphi.

The example $a=1$ is also instructive. The denominators $\Gamma(2-i)$
cause trouble for $i=2,3,...,n$. Hence one of the $k$'s may be chosen
freely, while all the others must satisfy $k_i+a-n+1=k_i-(n-2)\leq 0$.
The strict ordering of the $k$'s imposes the choice
\eqn\chimpos{ (k_1,k_2,...,k_{n-1})~=~(0,1,2,...,n-2) \qquad
{\rm and}\qquad k_n \geq n-1}
This results in an infinity of possible excitations $k_n=n-1+k$ above
the fundamental $k=0$. These contribute for
\eqn\contex{ Z_n(1,b)~=~\sum_{k=0}^\infty \left({t \over N}\right)^k
{n+k-1 \choose k} {\Gamma(k+b)\over \Gamma(b)} }
which is equivalent to \newphi.

The formula \order\ permits also to prove directly
the previously-mentioned $(a,b,n)$
symmetry of $Z_n(a,b)=Z_a(n,b)$ for $a=m$ any positive integer.
Indeed, for $a=1$, if we compute $Z_1(n,b)$ using \order, we find that
$k_1=k$ may take the fundamental value $k=0$ or any excitation
$k>0$, resulting in the same sum as \contex, therefore
\eqn\sumresul{Z_1(n,b)~=~Z_n(1,b) }
More generally, let us evaluate $Z_n(m,b)$ using \order.
The term $\Gamma(1+a-i)=\Gamma(m+1-i)$ in the denominator
causes problems for $i \geq m+1$ only. Assuming that $n>m$, this forces
$(n-m)$ of the $k$'s to satisfy $k_i+a-n+1=k_i-(n-m-1)\leq 0$.
These are necessarily
\eqn\fundax{(k_1,k_2,...,k_{n-m})~=~(0,1,...,n-m-1) }
as the sequence is strictly increasing.
The remaining $k$'s may be in either the fundamental state or any
excited state above it, namely
\eqn\excix{ n-m \leq k_{n-m+1} < k_{n-m+2}<...< k_n }
If we look instead at $Z_m(n,b)$, there are no bad denominators,
and all the excitations of the $m$ $k'$'s are permitted, namely
any
\eqn\anyx{ 0\leq k_1' < k_2'<...<k_m' }
The two sets of excitations \excix\ and \anyx\ can be mapped onto each
other by writing
\eqn\mapexit{ k_j~=~ n-m+k_{j+m-n}' \quad \forall\ j=n-m+1,n-m+2,...,n}
and it is a straightforward exercise to check that their contributions
in \order\ are identical.
This proves the symmetry $Z_n(m,b)=Z_m(n,b)$ for all $m$, $n$, $b$.

\subsec{The large $N$ limit: general solution}

When $N$ is large, with $n=zN$, we expect the sum \zndetf\ to be 
dominated by integers of the form $k_i=N \alpha_i$, where $\alpha_i$ 
are real numbers.  
The index $i$ itself ranging from $1$ to $n=zN$, we may let it scale as 
$i=Ns$, $s$ ranging from $0$ to $z$.
The sum \zndetf\
may therefore be approximated by the multiple integral
\eqn\intaprox{ Z_n(a,b)~\sim~ N^n\int_0^\infty d\alpha_1...d\alpha_n
e^{N^2 S(\alpha_1,...,\alpha_n;p,q,z)} }
where we have identified the effective action
\eqn\effac{\eqalign{ 
S(\alpha_1&,...,\alpha_n;p,q,z)~=~
{1\over N}\sum_{i=1}^n \big[
(\alpha_i+p-z)({\rm Log}(\alpha_i+p-z)-1)\cr
&+(\alpha_i+q-z)({\rm Log}(\alpha_i+q-z)-1)
-\alpha_i({\rm Log}(\alpha_i)-1) 
+\alpha_i\, {\rm Log}\, t\big] \cr
&+{1\over N^2}\sum_{1\leq i\neq j\leq n}
{\rm Log}|\alpha_i-\alpha_j|
-\int_0^z ds\big[(p-s)({\rm Log}(p-s)-1)\cr
&+(q-s)({\rm Log}(q-s)-1)
+s({\rm Log}(s)-1)+ s\, {\rm Log}\, t\big] \cr}}
by use of the Stirling formula.
The integral \intaprox\ is dominated by the saddle-point
$\partial S/\partial \alpha_i=0$, namely
\eqn\sapoeq{ 
{\rm Log}\left(t{(\alpha_i+p-z)(\alpha_i+q-z)\over \alpha_i}\right)~=~
{2\over N}\sum_{j\neq i} {1 \over \alpha_j-\alpha_i}}
for $i=1,2,...,n$. Introducing the large $N$ density
\eqn\densit{ \rho(\alpha)~=~\lim_{N\to \infty}
{1\over N}\sum_{i=1}^n \delta(\alpha-\alpha_i)}
with the normalization 
\eqn\noro{ \int d\alpha \rho(\alpha)~=~ z}
we get the limiting equation
\eqn\limeqa{\encadremath{ 
{\rm Log}\left(t{(\alpha+p-z)(\alpha+q-z)\over \alpha}\right)~=~
2\, P\int d\beta {\rho(\beta)\over \beta -\alpha} } }
where the symbol $P$ stands for the principal value.
As usual, we introduce the resolvent
\eqn\deresol{ \omega(\alpha)~=~\int d \beta {\rho(\beta)\over
\alpha-\beta} }
for any complex number $\alpha$. In particular, $\omega(\alpha)$
behaves for large $|\alpha|$ as
\eqn\behares{ \omega(\alpha)\sim {1\over \alpha}\int \rho(\beta)d \beta
~=~{z \over \alpha} }
by use of the normalization \noro .  Moreover, the saddle-point
equation \limeqa\ may be written as
\eqn\maywr{ \lim_{\epsilon \to 0^+}
\omega(\alpha+i\epsilon)+\omega(\alpha-i\epsilon)~=~-{\rm Log}\left(
t{(\alpha+p-z)(\alpha+q-z)\over \alpha}\right) }
If we can solve this equation, we can deduce the large $N$
value of the free energy from the effective action \effac.
Indeed, writing $S(\alpha_i^*;p,q,z;t)=f(p,q,z;t)$ the saddle-point
action, we may compute
\eqn\compmay{\eqalign{
t \partial_t f~&=~\sum_i \partial_{\alpha_i}S(\alpha_i)
t{d \alpha_i\over dt}\big\vert_{\alpha_i=\alpha_i^*}+t \partial_t S\cr
&=~ \int \beta \rho(\beta) d \beta -{z^2\over 2} \cr}}
where we have used \sapoeq\ at the saddle-point $\alpha_i=\alpha_i^*$
and the definition \densit\ of the limiting density $\rho$.
In terms of the resolvent
\eqn\resexo{\omega(\alpha)~=~ \sum_{i=1}^\infty{\Omega_i \over
\alpha^i}}
this simply reads
\eqn\simpre{ t \partial_t f(p,q,z;t)~=~ \Omega_2 -{z^2\over 2} }

To solve \maywr, let us perform the change of variables
\eqn\chvar{\eqalign{ \alpha~&=~z-r-2 \delta {\rm cosh}\, \phi \cr
p~&=~r+2 \delta {\rm cosh} \, \phi_1\cr
q~&=~r+2 \delta {\rm cosh} \, \phi_2\cr
z~&=~r+2 \delta {\rm cosh} \, \phi_3\cr}}
where $r$ and $\delta$ are two free parameters, to be fixed
later. We will take the variable $\phi=i\varphi$, 
$\varphi\in [0,\pi]$, to be purely 
imaginary on the support of the distribution $\rho$, so that
$z-r$ and $4\delta$ measure the center and width of this 
support.  We define 
\eqn\defth{\eqalign{ u=e^{-\phi},\qquad &T={\rm tanh}(\phi/2) =
{1-u\over 1+u} \cr
u_i~=~e^{-\phi_i},\qquad  &
T_i~=~{\rm tanh}(\phi_i/2)~=~{1-u_i \over 1+u_i}, \qquad i=1,2,3\cr}}
Then we have
\eqn\wehav{ {t\over \alpha}(\alpha+p-z)(\alpha+q-z)~=~
\delta t\, {4(T_1^2-T^2)(T_2^2-T^2)(1-T_3^2)\over
(1-T^2)(1-T_1^2)(1-T_2^2)(T_3^2-T^2)}}
Since on the support of $\rho$, $T=i\, $tan$(\varphi/2)$ satisfies 
$T^*=-T$, the saddle point equation \maywr\ has
the following solution 
\eqn\solresolvent{
\omega(\alpha)~=~-{\rm Log}\left( {2(T_1+T)(T_2+T)(1-T_3) \over
(1+T)(1+T_1)(1+T_2)(T-T_3)} \right) }
for the resolvent provided we take $\delta$ such that
\eqn\ifsat{ \delta t~=~{(1-T_1)(1-T_2)(1-T_3)\over
(1+T_1)(1+T_2)(1+T_3)}~=~u_1u_2u_3}
Moreover, the large $\alpha$ behavior \behares\ imposes that
\eqn\larbeha{\omega(\alpha)~\sim~u(u_1+u_2-{1\over u_3})~=~
-{zu\over \delta}}
namely that
\eqn\nameth{ u_1u_2u_3({1\over u_3}-u_1-u_2)~=~zt}
Using the third line of \chvar , this fixes the value of $r$ to 
\eqn\alav{ r~=~-\delta (u_1+ u_2 + u_3)}
Substituting this value into the second and third lines of \chvar, and
using the value \ifsat\ for $\delta$, we get
\eqn\getr{\eqalign{
u_1u_2u_3({1\over u_1}-u_2-u_3)~&=~pt\cr
u_1u_2u_3({1\over u_2}-u_1-u_3)~&=~qt\cr}}
Introducing the new variables
\eqn\nwva{ U_1=u_2u_3 \qquad U_2=u_1u_3 \qquad U_3=u_1u_2}
we may recast \nameth\ and \getr\ into
\eqn\generp{\eqalign{
U_1(1-U_2-U_3)~&=~pt\cr
U_2(1-U_1-U_3)~&=~qt\cr
U_3(1-U_1-U_2)~&=~zt\cr}}
By eliminating $U_2$ and $U_3$ as
\eqn\othsol{
\eqalign{ U_2~&=~ {1 \over 2}
(1-{pt \over U_1}+{(q-z)t\over 1-U_1})\cr
U_3~&=~{1 \over 2}(1-{pt \over U_1}
-{(q-z)t\over 1-U_1})\cr}}
this reduces to the following
5-th order equation for $U_1\equiv U_1(p,q,z;t)$
\eqn\sixo{\encadremath{
U_1^2(1-U_1)^2(1-2U_1+2(p-q-z)t)~
=~t^2\big((1-U_1)^2 p^2-U_1^2(z-q)^2\big)}}
We have to retain the unique solution behaving for small $t$
as $U_1\sim pt$ (this is the unique small $t$ behavior compatible
with Eq.(4.49) below). The other $U$'s are obtained
by exchanging the values of $p,q,z$, namely
\eqn\othU{\eqalign{
U_2(p,q,z;t)~&=~U_1(q,p,z;t)\cr
U_3(p,q,z;t)~&=~U_1(z,q,p;t)\cr}}
We then have to plug these solutions back into
\eqn\plug{\eqalign{
\delta~&=~{1\over t} \sqrt{U_1U_2U_3} \cr
r~&=~-{1\over t} (U_1U_2+U_2U_3+U_1U_3) \cr
u_i~&=~ {\sqrt{U_1U_2U_3} \over U_i} \qquad i=1,2,3\cr}}

For instance, the density associated to the resolvent \solresolvent\ reads
\eqn\readro{\eqalign{
\rho(\alpha)~&=~{1\over 2i\pi}\lim_{\epsilon\to 0^+}
\omega(\alpha+i\epsilon)-\omega(\alpha-i\epsilon) \cr
&=~{1 \over 2 \pi} \big( \varphi-2 (\Psi_1+\Psi_2+\Psi_3) \big)\cr}}
with $\alpha=z-r-2 \delta {\rm cos}(\varphi)$,
and we have introduced the quantities
\eqn\introq{ \Psi_i~=~{\rm tan}^{-1}
\big({{\rm tan}({\varphi\over 2})\over T_i}\big) }

The free energy is easily obtained from \simpre\ by expanding
the resolvent \solresolvent\ up to the second order in $1/\alpha$.
We find
\eqn\finF{\encadremath{t\partial_t f(p,q,z;t)~=~\Omega_2-{z^2\over 2}~=~
{U_1 U_2U_3\over t^2} (1-U_1-U_2-U_3)} }
Note that this is explicitly symmetric in $p,q,z$ as expected.

\subsec{Particular cases}

In the passage \othsol\ from \generp\ to \sixo , we implicitly assumed
that $U_1$ was different from $0$ and $1$. Let us examine these two cases
in more details.

If $U_1=0$, we read from \generp\ that $p=0$. Conversely, if $p=0$,
we deduce from \generp\ that either $U_1=0$ identically or $U_2+U_3=1$ 
identically. The latter solution is however not compatible with the 
requirement that $U_2+U_3\sim (q+z)t$ at small $t$ and should be discarded.
Setting $U_1=0$ in \generp\ leads to
\eqn\uundeux{\eqalign{U_2-U_2U_3&=qt \cr U_3-U_2U_3&=zt\cr}}
Since $\delta=0$ in this limit, the support of the distribution $\rho$
reduces to one point located at
\eqn\concsup{\alpha={U_2U_3\over t}}
which, using \uundeux , is solution of
\eqn\solastar{t(\alpha+q)(\alpha+z)=\alpha}
We thus recover the saddle-point result \sappo\ with $p\leftrightarrow z$.

If $U_1=1$, we read from \generp\ that $q=z$. Conversely, if $q=z$,
we deduce from \generp\ that either $U_1=1$ identically or $U_2=U_3$
identically. From the requirement that $U_1\sim pt$ for small $t$, 
we see that it is now the latter solution which has to be kept while
the solution $U_1=1$ is simply unphysical.
Let us nevertheless study this particular case $q=z$,
now described by its correct solution $U_2=U_3$.
We now get a third order equation for $U_1$
\eqn\thiro{Q(U_1,t)~=~ U_1^2(1-2U_1+2(p-2z)t)-p^2t^2~=~0}
This fixes $U_1$ as a function of $t$. Writing $\partial_tU_1
=-\partial_tQ(U_1,t)/\partial_uQ(U_1,t)$,
we find a first order critical point $t=t_*(p,z,z)$, $U_1=u_*$, 
whenever $\partial_u Q(u_*,t_*)=0$, hence
\eqn\secordo{\partial_u Q(u_*,t_*)~=~ 2u_*(1-2u_*+2(p-2z)t_*)-2u_*^2~=~0}
By combining with \thiro , we get $u_*^3=(pt_*)^2$.
As $u_*>0$, it is convenient to introduce the parametrization
\eqn\conpara{ u_* ~=~{1\over 4 \cos^2 {\theta\over 3}} }
where
\eqn\ranges{\eqalign{&\theta\ge 0\ \ {\rm when} \ \
{1\over 4}\leq u_*  \cr
\theta~&=~i\varphi,\  \varphi\ge 0\ \ {\rm  when}\ \ 0\leq u_* \leq
{1\over 4}\cr}}
The critical point is then
\eqn\criqz{ t_*(p,z,z)~=~{\epsilon \over 8 p \cos^3 {\theta\over 3} } }
with $\epsilon=\pm 1$.
The values of $\theta$ and $\epsilon$ are fixed by the ratio
\eqn\thenrati{\eqalign{ {z\over p}~=~{q \over p}~&=~
{1+\epsilon \cos \theta\over 2}\cr
&=~\left\{ \matrix{ \cos^2 {\theta \over 2} & \ {\rm if}\ &
\epsilon=1\cr
\sin^2 {\theta\over 2} & \ {\rm if}\ & \epsilon=-1 \cr} \right.\cr} }
For $z/p> 1$, \thenrati\ has exactly one solution corresponding to
$\epsilon=1$, $z/p=\cosh^2{(\varphi/2)}$, and we find a critical point
at $t_*=1/(8p\cosh^3{(\varphi/3)})$. We thus have
\eqn\valtstar{t_*(p,z,z)={1\over 8p\cosh^3\left({2\over 3} {\rm arccosh}
\sqrt{z\over p}\right)}}
For $z/p<0$, \thenrati\ gives
also exactly one solution $(u_*,t_*)$ with now $\epsilon=-1$. 
However, this point is not on the correct branch of the solution 
of \thiro , i.e. that satisfying $U_1\sim pt$ at small $t$. 
This is easily checked by solving explicitly \thiro, which is 
quadratic in $t$ in the form $t(U_1)$ rather than $U_1(t)$, 
and choosing the solution with the correct small $t$ behavior.
We thus deduce that there is no critical point ($R=+\infty$) 
for $z/p<0$.

For $0\le z/p \le 1$ \thiro\ has an infinite number of 
solutions of the form $\epsilon=1$, $\theta=\theta_0+2k\pi$, 
$\theta=(2k+1)\pi -\theta_0$ and $\epsilon =-1$, 
$\theta=\theta_0+(2k+1)\pi$, $\theta=(2k+2)\pi-\theta_0$
with $\theta_0\in [0,\pi[$ and $k$ a non negative integer. 
This leads to exactly three candidates
for the critical point $t_*=1/(8p\cos^3(\theta_0/3))$, 
$t_*=1/(8p\cos^3(\theta_0/3+2\pi/3))$ and 
$t_*=1/(8p\cos^3(\theta_0/3+4\pi/3))$. 
Here again, it is easy to check that the critical point attached 
to the correct branch is the first one, namely 
\eqn\waltstar{t_*(p,z,z)={1\over 8p\cos^3\left({2\over 3} {\rm arccos}
\sqrt{z\over p}\right)}}
where the arccos is taken in the range $[0,\pi[$. 

When $p=q=z$, we recover the critical value $t_*(z,z,z)=1/8z$ of \tstarzzz\ 
by setting $\theta_0=0$, with moreover $u_*=1/4$.
This case $p=q=z$ can be solved more directly by noting that,
from \generp , we then have $U_1=U_2=U_3\equiv U$, where
$U$ is the solution of a quadratic equation $U(1-2U)=zt$, namely
\eqn\Ueq{U~=~{1\over 4}(1-\sqrt{1-8zt})~=~zt C(2zt)}
The corresponding free energy reads
\eqn\frepqeg{t \partial_t F(z;t)~=~{U^3\over t^2}(1-3U)~=~
{z\over 8t}\big((8zt-1)C(2zt)+1-6zt\big)}
which coincides with our previous result \markedge\-\nuk.

\subsec{Critical Points}

The solution $U_1$ to \sixo\ such that $U_1\sim pt$ when $t\to 0$
will in general have a convergent
series expansion until we reach a critical point $t=t_*(p,q,z)$,
corresponding to the convergence radius $R=t_*$. To determine
this point, we write the (finite) Taylor series of the polynomial
\eqn\ftayl{ P(u,t)~=~u^2(1-u)^2(1-2u+2(p-q-z)t)-(1-u)^2p^2 t^2
+u^2 (q-z)^2 t^2 }
namely
\eqn\talo{\eqalign{ P(u,t)~&=~P(u_*,t_*)+(u-u_*)\partial_uP(u_*,t_*)
+{(u-u_*)^2\over 2}\partial_u^2P(u_*,t_*)\cr
&+(t-t_*)\partial_tP(u_*,t_*)+...\cr} }
Imposing that $P$ vanishes, we get a first order critical point
\eqn\firtord{(u_*-u)~\sim~ (t_*-t)^{1/2}}
by demanding that
\eqn\critipo{P(u_*,t_*)~=~\partial_uP(u_*,t_*)~=~0}
while
\eqn\satisur{\partial_u^2P(u_*,t_*)\neq 0\qquad
\partial_t P(u_*,t_*)\neq 0}
This means that the polynomial $P(u,t_*)$ takes the form
\eqn\condip{P(u,t_*)~=~(u-u_*)^2(-2u^3+a u^2 + b u + c)}
for some numbers $a,b,c$ to be determined.
Introducing the parameters
\eqn\intopa{\alpha=2(p-q-z)t_*\qquad \beta=p^2 t_*^2 \qquad
\gamma=(q-z)^2 t_*^2}
and identifying \ftayl\ with \condip, we get
\eqn\getwe{ \eqalign{
a+4 u_*~&=~\alpha+5\cr
b-2 a u_*-2 u_*^2~&=~-2(\alpha+2)\cr
c-2bu_*+a u_*^2~&=~1+\alpha-\beta+\gamma\cr
b u_*^2 -2 c u_*~&=~ 2 \beta\cr
c u_*^2~&=~ - \beta\cr}}
Eliminating $\beta$ from the last two lines of \getwe, and
$\alpha$ from the first two, we find
\eqn\findwe{\eqalign{
b~&=~ 2 c {1-u_*\over u_*}\cr
a~&=~ 3 -u_*-{c \over u_*} \cr} }
Here we can assume that $u_*\neq 0,1$, since the situation where $U_1=0$ 
or $1$ has been treated in the previous section. 
The equations \findwe\ translate into the following
parametrization of $\alpha$, $\beta$, $\gamma$ of \intopa\
\eqn\neqp{ \eqalign{
\alpha~&=~2(p-q-z)t_*~=~ 3 u_*-2 +\xi \cr
\beta~&=~p^2t_*^2~=~\xi u_*^3\cr
\gamma~&=~(q-z)^2 t_*^2~=~(1-u_*)^3 (1-\xi)\cr}}
where we have set $c=-\xi u_*$.
This may be viewed as a parametric curve $(pt_*,qt_*,zt_*)=h(u_*,\xi)$,
where $u_*$ and $\xi$ are real parameters, with $u_*\xi\geq 0$ and
$(1-u_*)(1-\xi)\geq 0$. This in turn permits to express for instance
\eqn\tsart{\encadremath{t_*~=~{\epsilon \over p}\sqrt{\xi u_*^3}}}
where $\epsilon=$sgn$(pt_*)$,
as an implicit function of the ratios $q/p$ and $z/p$, where
\eqn\ratios{\encadremath{\eqalign{
{q\over p}~&=~{1\over 2}\big( 1+\epsilon' {1-u_*\over u_*}
\sqrt{(1-u_*)(1-\xi)\over u_*\xi}-\epsilon 
{3u_*-2+\xi\over 2u_*\sqrt{\xi u_*}}\big)\cr
{z\over p}~&=~{1\over 2}\big( 1-\epsilon' {1-u_*\over u_*}
\sqrt{(1-u_*)(1-\xi)\over u_*\xi}-\epsilon 
{3u_*-2+\xi\over 2u_*\sqrt{\xi u_*}}\big)\cr } } }
and where $\epsilon'=$sgn$(p(q-z))$.
Note that the transformation $(\xi,u_*)\to (1-\xi,1-u_*)$
simply amounts to
\eqn\transfo{\matrix{p & \to & \epsilon'(q-z) \cr
q & \to &
{1+\epsilon'\over 2} p +{\epsilon'-1\over 2} q
-\epsilon' z\cr
z & \to &
{1-\epsilon'\over 2} p +{\epsilon'-1\over 2} q
-\epsilon' z\cr} }

We must now check the validity of this critical point, by making
sure that \satisur\ is satisfied. At the critical point, we find
\eqn\cirtfin{\eqalign{
\partial_tP(u_*,t_*)~&=~ {1\over t_*}u_*^2 (1-u_*)^2 (u_*-\xi)  \cr
\partial_u^2P(u_*,t_*)~&=~ 6 u_* (1-u_*)(u_*-\xi) \cr}}
Provided we take $u_*\neq 0,1,\xi$, we see that \satisur\ will hold.
As mentioned above, the case $u_*=0$ corresponds to $p=0$ 
and has already been studied in the
previous section. Taking $u_*=1$ implies that $q=z$, a particular case 
also treated in the previous section. Note that in this case, the 
correct solution for the critical point is not at $u_*=1$, 
which lies on a wrong branch, but 
at some $u_*<1$ given by \conpara. The case $u_*=\xi$ corresponds 
to symmetric counterparts
of these particular cases. Indeed, from \neqp, we find that
$pt_*=\epsilon u_*^2$, $(q-z)t_*=\epsilon\epsilon'(1-u_*)^2$
and $(q+z)t_*=\epsilon u_*^2+1-2 u_*$, hence
\eqn\solmut{(qt_*,zt_*)~=~
((1-u_*)^2, 0),\ \  (0,(1-u_*)^2),\ \  (1-2u_*,-u_*^2),\ \
(-u_*^2,1-2u_*)}
according to the values of the signs
$(\epsilon,\epsilon')=(+,+),(+,-),(-,+),(-,-)$.
The first two cases correspond to $z=0$ or $q=0$ and are equivalent
to the $p=0$ particular case under the change $p\leftrightarrow z$
or $p\leftrightarrow q$. 
The two last cases correspond to either $p=q$ or $p=z$, which are
now equivalent to the $q=z$ solution. Here again, $u_*=\xi$ lies
on a wrong branch of the solution.  

From now on, we shall ignore the particular cases by 
demanding that none of the parameters $p$, $q$ and $z$ vanishes
and that they are all different. Using the $(p,q,z)$ symmetry,
we can furthermore assume without loss of generality 
that $-1\le z/p <q/p <1$, with in particular $\epsilon'=1$.
For fixed $q/p$ and $z/p$ in this range, any solution of 
\ratios\ is a candidate to describe a first order critical point,
of the form \firtord, with $t_*$ given by \tsart.
We still have to make sure that both terms in \satisur\ have the
{\it same sign}, in order for the solution to read
$u_*-u\sim (1-t/t_*)^{1/2}$. This is the case if and only if
\eqn\signcon{ 0~<~u_*~<~1 }
which in turn implies $0<\xi<1$.
The parametrization \ratios\ is further restricted 
by demanding that the critical value
$u_*$ indeed corresponds to a solution $U_1(t)=pt+O(t^2)$ for small $t$.
Using the equation 
\sixo, we may solve for $t$ as a function of $U_1$. We have to pick the
right branch of this quadratic equation, namely that for which
$t(U_1)={U_1\over p}+O(U_1^2)$ for small $U_1$.
It reads
\eqn\ribran{\eqalign{ 
pt(u)~&=~{u(1-u) 
\over (1-u)^2-\left({q-z\over p}\right)^2 u^2} \times \Bigg[ 
\left(1-{q+z\over p}\right)u(1-u)\cr
&+\sqrt{u^2(1-u)^2
\left(1-{q+z\over p}\right)^2+(1-2u)\left((1-u)^2-\left(
{q-z\over p}\right)^2u^2\right)} \Bigg]}}
In this formulation, a critical point corresponds to a local 
extremum of $pt(u)$.
The parametrization \ratios\ corresponds to this branch provided
\eqn\contio{ \epsilon~=~{\rm sgn}(\xi-u_*) }
This is obtained by substituting \ratios\ into \ribran, and 
comparing the result with \tsart. 
So finally, we have fixed the values of $\epsilon$ and
$\epsilon'$ in \ratios.
To ensure the uniqueness of the critical point we found, we
have to check that the parametrization \ratios\ is bijective.
We must therefore compute the Jacobian $J$ of the map $(u_*,\xi)\to 
(q/z,p/z)$ induced by \ratios: we find
\eqn\jacomap{ J~=~{3\over 16 u^4 \xi^2} 
\epsilon (u-\xi)^2\sqrt{1-u \over 1-\xi} }
This is non-vanishing for $0< u \neq \xi < 1$.
We therefore have for fixed $q/p$ and $z/p$ at most two
acceptable solutions of \ratios, one in the range
$0<u_*<\xi<1$ with $\epsilon=1$ and one in the
range $0<\xi<u_*<1$ with $\epsilon=-1$.
This latter solution can be eliminated by noting that
it corresponds to a value of $t_*$ given by \tsart\ 
such that $pt_*<0$, which is unphysical. Indeed, starting 
from $pt\sim U_1$ at small $t$, one cannot reach continuously
an extremum $(u_*,t_*)$ of $t(u)$ with $u_*>0$ and $pt_*<0$ 
without first reaching an extremum at some $u_*>0$, $pt_*>0$.  
This fixes $\epsilon=1$ and $0<u_*<\xi<1$, which leaves
us to at most one solution of \ratios. In this sense,
the parametrization \tsart\-\ratios\ leads to a unique expression
$t_*(p,q,z)=(1/p)T(q/p,z/p)$ for those values of $q/z$ and $z/p$
in the range $-1\le z/p < q/p <1$ for which a critical point
actually exists. 

To determine which values of $q/p$ and $z/p$ give rise to a critical
point, we distinguish two situations in the formula \ribran, as
described in more details in Appendix B.  For $z/p<0$, the 
discriminant in \ribran\ does not vanish. The function $pt(u)$
is a strictly increasing function from $t=0$ at $u=0$ to $t=+\infty$
at the value $u=1/(1+(q-z)/p)$ where the denominator vanishes.
There is no critical point in this case, hence $R=+\infty$.
For $z/p>0$, the discriminant in \ribran\ vanishes at some value
$u_0$ of $u$ strictly less than the limiting value $1/(1+(q-z)/p)$ for
the vanishing of the denominator. At $u_0$, the slope of $t(u)$
is $-\infty$, which implies the existence of a critical point at
some $u_*$ between $0$ and $u_0$ with $pt_*>0$.

In conclusion, we find a critical point only if $0<z/p<q/p<1$,
i.e. if $p$, $q$ and $z$ have the same sign, which in turn is also
the sign of $t_*$.

\fig{Critical value $t_*(p,q,z)$ in the domain $-1\le z/p \le q/p \le 1$.
A critical point is reached only in the sub-domain $z/p\ge 0$ (shaded
region). The limiting values of $t_*$ are explicited at the boundary
of this sub-domain.
}{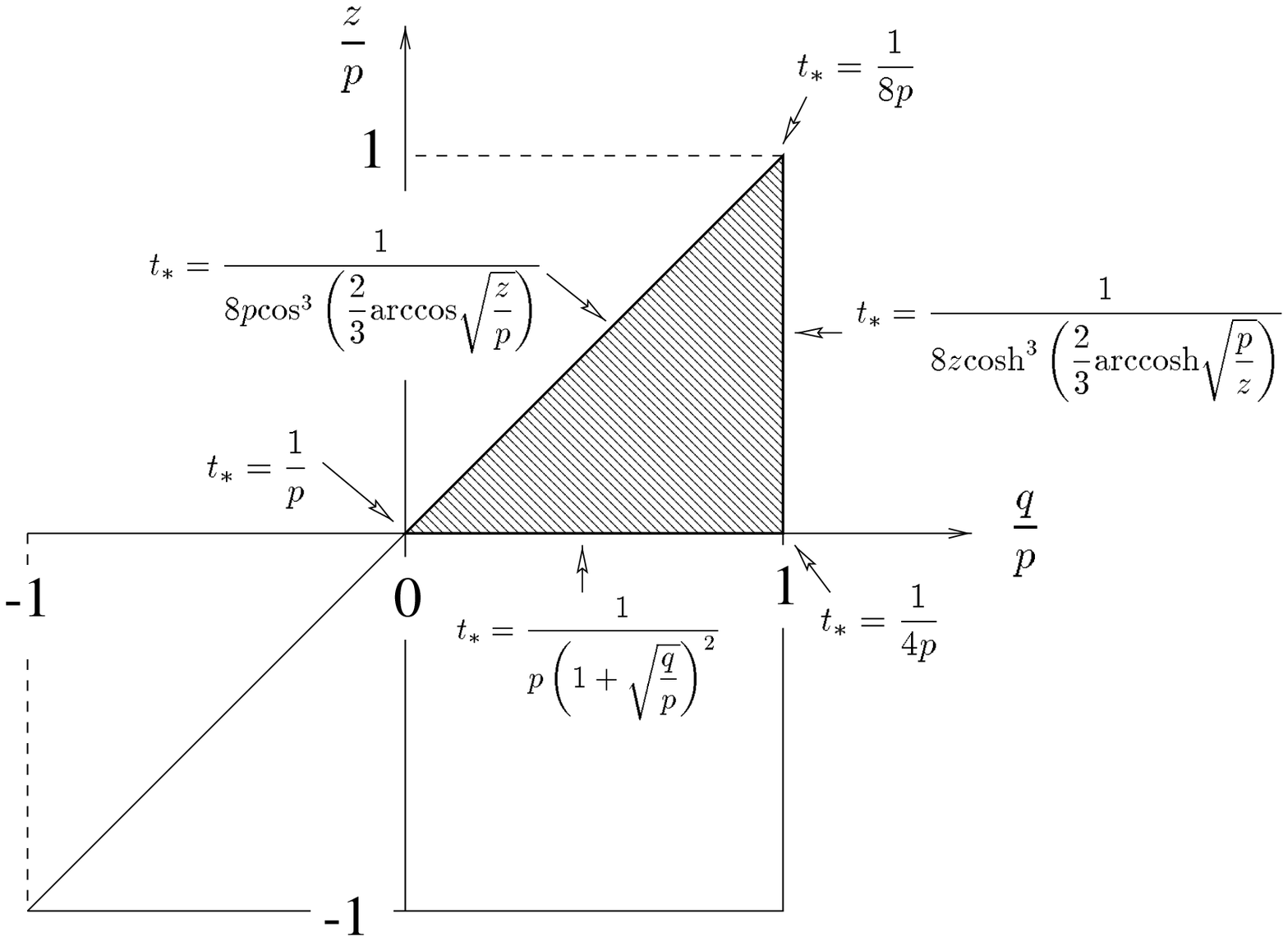}{12.cm}
\figlabel\critic
Our results are summarized in Fig.\critic, where we have extended the domain
to $-1\le z/p \le q/p \le 1$ by reintroducing the results of the
previous section for the particular cases.

Using the results \cirtfin, we may finally compute the first
sub-leading coefficient in the expansion
\eqn\Uexpo{ U_1~=~ u_*- \kappa (t_*-t)^{1\over 2} + ...}
namely
\eqn\firstco{ \kappa~=~ \sqrt{u_*(1-u_*)\over 3 t_*} }

The free energy follows from \finF. Actually, it is simpler
to compute
\eqn\finfgt{ t^2 (t\partial_t)^2 f~=~ U_1U_2U_3}
easily obtained by use of \generp.
Substituting the values \othsol\ for $U_2$ and $U_3$, this reads
\eqn\redsthi{\eqalign{
t^2 (t\partial_t)^2 f~&=~ {U_1\over 4}\big( (1-{pt\over U_1})^2-
({(q-z)t\over 1-U_1})^2\big)\cr
&=~{1\over 2}\big( (1+(p-q-z)t)U_1-pt-U_1^2\big)\cr}}
Substituting the expansion \Uexpo, we get the expansion
\eqn\sinfgu{\encadremath{ t^2(t\partial_t)^2f~=~
{1\over 4} u_* (\sqrt{\xi}-\sqrt{u_*})^2 +
{u_*-\xi\over 4}  \sqrt{u_*(1-u_*)
\over 3} \sqrt{1-{t\over t_*}} + ... } }
with a non-vanishing singular term if $u_*\neq0,1,\xi$.
In particular, for $p=q=z$, we have $u_*=1/4$ and 
$\xi=1$ so that \sinfgu\ yields
\eqn\twoixz{ t^2 (t\partial_t)^2 f(z,z,z,t)~=~ {1\over 64}-{3\over 64}
\sqrt{1-8tz} + ... }
in perfect agreement with \provid\-\viprod.

The above result \sinfgu\ leads to the
string susceptibility exponent $\gamma_{str}=-1/2$,
governing the critical behavior of the free energy
$f\sim (t_*-t)^{2-\gamma_{str}}$.  This coincides with the string
susceptibility exponent of pure gravity, attached to the plain
enumeration of planar diagrams.
This universal result holds for all the values of parameters $(p,q,z)$ 
where a critical point is reached, except in the particular cases where
one of these parameters is exactly zero, in which case we find
the behavior \tapproa .
The convergence radius $R=t_*$ may be related to the entropy $s$
of three-coloring of the vertices of random triangulations, namely
$R=e^s$. Actually, as a consequence of
\sinfgu, the contribution to the free energy with a
fixed number of black-white
edges $n$ ($=1/2\times$ the number of triangular faces)
behaves for large $n$ as
\eqn\behnla{ \omega_n(p,q,z)~\sim~ {e^{ns(p,q,z)}\over n^{7\over 2}}}
where the three-coloring entropy per black-white edge reads
\eqn\entrop{ s(p,q,z)~=~-{\rm Log}\, t_*(p,q,z) }
with $t_*$ as in \tsart.

\subsec{Double Scaling Limit}

We may now reconsider the all-genus expansion of the free energy
\foropo\
\eqn\fregty{f(p,q,z;t;N)~=~f_n(p,q,1/t,N)~
=~\sum_{h\geq 0} N^{2-2h} \varphi_h(p,q,z;t) }
where the genus $h$ free energy $\varphi_h$ is a quasi-homogeneous
function, satisfying
$\varphi_h(\lambda p,\lambda q,\lambda z; t/\lambda)=\lambda^{2-2h}
\varphi_h(p,q,z;t)$.
When $t$ approaches the critical point $t_*(p,q,z)$ of \tsart, the
leading singular part of $\varphi_0$ behaves as $(t_*-t)^{5/2}$,
and has a prefactor $N^2$.
This suggests to perform the double-scaling limit $t\to t_*$,
$N\to \infty$, by keeping the parameter
\eqn\paray{ y~=~ N^{4/5} \, {t_*-t\over t_*} }
fixed. Let us use the expansion parameter $\epsilon=1/N$.
At the vicinity of $t_*$, let us write the total free energy
\eqn\totgref{\eqalign{ f(p,q,z;t;N)~&=~ \epsilon^{-2} g_0(p,q,z) +
\epsilon^{-6/5} g_1(p,q,z) y +\epsilon^{-2/5} g_2(p,q,z) y^2 \cr
&+ \epsilon^0 f(y;p,q,z)
+\epsilon^{2/5} g_3(p,q,z) y^3 +O(\epsilon^{4/5})\cr}}
This defines the scaling function $f(y;p,q,z)$, capturing the
contributions
to the free energy from all genera in  the double scaling limit
\paray. Its genus zero ($y\to \infty$) contribution reads
\eqn\genzerf{ f(y;p,q,z)~=~ {u_*-\xi\over 4}
\sqrt{u_*(1-u_*)\over 3} \, y^{5\over 2} +...}
as a consequence of \sinfgu.
To derive an equation for $f(y;p,q,z)$, let us start from the recursion
relation \satiphi, and expand it in powers of $\epsilon$.
With $f_n={\rm Log}\, Z_n=f(p,q,z;t;N)\equiv f(z,t)$, we find
\eqn\satiexpo{\eqalign{ f(z+\epsilon,t)+f(z-\epsilon,t)-2f(z,t)~&=~
{\rm Log} \big( 1+\epsilon^{-2}{t\over z} (t\partial_t)^2 f(z,t)\big)\cr
&=~\epsilon^2 \partial_z^2f +{1\over 12} \epsilon^4 \partial_z^4 f+...
\cr}}
Substituting the expansion \totgref\ for $f$, we finally get at the
order
$\epsilon^{4/5}$ the equation
\eqn\plone{\encadremath{ 
{1\over 6} \partial_y^4 f +(\partial_y^2 f)^2 ~=~ k^2 y }}
where $k$ is a function of $p,q,z$ only, fixed by the genus zero
behavior \genzerf\ to be
\eqn\valaopp{ k~=~{5\over 4} (u_*-\xi)\sqrt{3 u_*(1-u_*)} }
Eq.\plone\ is nothing but the Painlev\'e I equation of two-dimensional
pure quantum gravity, for the string susceptibility $\partial_y^2f$.
The dependence on the parameters $p,q,z$ is entirely contained in
the function $k$. Note that $k$ is symmetric in $p,q,z$, as
\eqn\kis{ k^2~=~75 u_*^{(1)} u_*^{(2)} u_*^{(3)}
(1-u_*^{(1)})(1-u_*^{(2)})(1-u_*^{(3)}) }
where $u_*^{(1)}=u_*(p,q,z)$, $u_*^{(2)}=u_*(q,p,z)$,
$u_*^{(3)}=u_*(z,q,p)$ denote
respectively the critical values of the functions $U_1$, $U_2$, $U_3$
of \generp.
(Recall that we have written $u_*$ as an implicit function of $p,q,z$
through the parametrizations \ratios, which must be inverted.)

We conclude that although our model involves more than the mere
enumeration of random surfaces, its double-scaling limit satisfies
the same universal Painlev\'e I equation as pure gravity, up to
a simple rescaling involving the function $k(p,q,z)$.

\newsec{Conclusion}

In this paper, we have investigated the partition function \expgrt\
of the vertex tri-coloring problem for random triangulations of
arbitrary genus.  In addition to the exact evaluation of the
tri-coloring entropy \entrop\-\tsart, 
we have shown that the model lies
in the universality class of pure two-dimensional quantum gravity,
with a string susceptibility exponent $\gamma_{str}=-1/2$.

\subsec{Folding of random triangulations}

It is interesting to note that, in the vertex tri-coloring problem,
the coloring itself does not give rise to any degree
of freedom since for a given triangulation, there is only one
way to color its vertices (up to global permutations of
the three colors). Varying $p$, $q$ and $z$ only amounts to favoring
one of these global permutations. This situation is to be contrasted 
with the tri-coloring of the edges of triangulations. There, even for 
the regular triangular lattice, the number of tri-colorings increases 
exponentially with the system size.

Therefore, in our problem of vertex tri-coloring, all the entropy
comes from the sum over triangulations itself, as for pure gravity. 
The novelty here comes from the fact that this sum is restricted 
to {\it tri-colorable triangulations}. In particular, this tri-colorability 
constraint implies that all vertices have an even number of surrounding 
triangles. This requirement is clear since the vertices around a vertex 
of a given color are bi-colored with alternating 
colors. Conversely, this parity condition turns out to be a sufficient 
condition to ensure tri-colorability when applied to triangulations 
with genus zero, as discussed in \SM . This is not the case however 
for triangulations with higher genus.  
For $p=q=z$ where we do not distinguish between the
different colors, we have found that the sum over connected 
planar tri-colorable triangulations obeys a quadratic equation
and can be given in terms of the generating function for 
Catalan numbers. This result is somewhat surprising since Catalan
numbers are usually associated with the enumeration of objects which 
can be simply decomposed into two similar smaller objects 
(like trees, arches, parentheses). We do not see any such decomposition
here.

In is interesting to reinterpret our results in the language
of folding of the triangulations. If we imagine a triangulation
made of equilateral triangles with unit side length, we may want to map 
its vertices onto those of the regular triangular lattice in the plane, 
with positions $a\vec e_1+b\vec e_2+ c\vec e_3$, ($a$, $b$ and $c$ integers)
where $\{\vec e_i\}$ is a set of three unit vectors in the plane 
satisfying $\vec e_1+ \vec e_2 +\vec e_3=\vec 0$.
Such a mapping will be called a folding of the triangulation if it sends
nearest neighbors of the triangulation onto nearest neighbors on the lattice,
thus preserving the equilateral nature of the triangles.
Two neighboring triangles will be either side by side in the plane
or on top of each other with a fold between them.
All triangulations cannot be folded onto the regular planar
triangular lattice. The possibility for a triangulation to be folded
is precisely equivalent to the tri-colorability of its vertices. 
To see this equivalence, we note that the regular triangular lattice can
be decomposed into three regular triangular sublattices such that any
neighboring vertices belong to two different sublattices. If we color
each of these sublattices with three different colors, this coloring
induces a tri-coloring of the vertices of any folded triangulation,
which is thus tri-colorable. Conversely, if the triangulation is tri-colored,
it can clearly be mapped onto a single triangle, by sending all vertices
of the same color onto the same vertex of the triangle.
We see here that the tri-coloring of the vertices can be viewed as a {\it particular}
folded configuration of the triangulation where all triangles are
mapped onto the same triangle. We call this configuration the {\it complete
folding} since all edges of the lattices sustain a fold. Our result
simply counts completely folded random triangulations with a different 
weight to the three different image
vertices. Clearly, the complete folding is only a particular folded state
and there are in general many different ways to fold a given tri-colorable 
triangulation, with an image covering in general several triangles in the plane.
This corresponds to an additional degree of freedom which is nothing but 
the tri-coloring of the edges of the triangulation. Indeed, any edge of the 
triangulation is sent onto one of the three vectors 
$\vec e_1$, $\vec e_2$ or $\vec e_3$ and
we can assign a color to this edge accordingly. In conclusion, the problem of
folding of random triangulations can be identified with that of tri-coloring
of {\it both} its vertices and its edges, a problem yet to be solved. 
 
\subsec{Hirota equation for multi-matrix models}

The general solutions we found take the
form of either some recursion relations \satiphi, \hiroz\ or
an explicit multiple sum over strictly increasing sequences of integers
\order. The latter is a discrete analogue
of the initial matrix integral itself, as it involves the
square of the Vandermonde determinant of the integers summed over.
Actually the large $N$ limit of this sum is dominated by the integral
\intaprox, which resembles the large $N$ approximation to the
$n\times n$ {\it one-matrix} integral
\eqn\onemapro{ \int dM \, e^{-N\, U(M;p,q,z)} }
where the potential $U$ is derived from the effective action $S$ of
\effac, namely $U(M;p,q,z)=N S(\lambda_1,...,\lambda_n;p,q,z)$,
where $\lambda_i$ denote the eigenvalues of $M$.  This resemblance is
however not quite an identity, as the corresponding eigenvalue integral
should extend over the whole real line for \onemapro, whereas it extends
only over $[0,\infty[$ in \intaprox.

The existence of a discrete Hirota bilinear equation for $Z_n$ is
actually generic for the following multi-matrix integral
\eqn\multi{ Z~=~\int dM_1 dM_2...dM_k \,
\det(1-M_1)^{-a} \det(1-M_k)^{-b} \, e^{-N{\rm Tr} V(M_1,...,M_k)}}
over $n\times n$ Hermitian matrices, with the potential
\eqn\potmulti{
V(M_1,...,M_k)~=~\sum_{i=1}^k V_i(M_i) +\sum_{j=1}^{k-1} u_j M_jM_{j+1}}
with arbitrary potentials $V_i$ (including possible logarithmic pieces).
Indeed, using the Itzykson-Zuber integral \iniz, we may reduce
$Z$ to an integral over the eigenvalues $m^{(i)}_j$ of the $M_i$'s.
Proceeding exactly as in Sect.4.1, we may recast \multi\ into
the determinant
\eqn\recamult{ \phi_n~=~\det\left[ \int d x_1d x_2...dx_k
(1-x_1)^{i-a-1} (1-x_k)^{j-b-1} e^{-NV(x_1,...,x_k)}
\right]_{1\leq i,j\leq n} }
up to an unimportant multiplicative constant.
This is clearly a solution of the Hirota equation \hiro. This solution
is singled out by its initial value $\phi_1$, which is a function of
the details of the potential $V$ (and in particular of the parameters
$u_1$, $u_2$,...$u_{k-1}$).
The case $k=1$ of a one-matrix model is also instructive. In that case,
we find that \KAZ\
\eqn\onedetphi{\phi_n~=~\det\left[\int dm (1-m)^{i+j-a-2} \, e^{-NV(m)}
\right]_{1\leq i,j\leq n} }
satisfies the following bilinear equation
\eqn\bilip{\encadremath{
\phi_{n+1}(a+2)\phi_{n-1}(a)~=~\phi_n(a+2)\phi_n(a)
-\big(\phi_n(a+1)\big)^2 } }

In this paper, we have seen however the limits of the use of the
Hirota equation, insofar as the large $N$ limit, for instance,
is concerned.
Because the Hirota equation
is only a multi-dimensional recursion relation, strongly
relying on its initial conditions, it makes
it quite difficult to extract any asymptotic result. We have had to
resort to a different approach to reach that goal.  Perhaps a suitable
mixing of the orthogonal polynomial solution and of the Hirota equation
could lead to some more definite results: this is still an open
question.

Finally, let us mention that our solution has displayed
striking connections to the arches and meanders enumeration
problems and their
reexpressions within the framework of the Temperley-Lieb algebra
\DGG\-\DGGB. We intend to return to these aspects in a later
publication.

\bigskip

\noindent{\bf Acknowledgements}

We thank V. Kazakov and P. Wiegmann for illuminating discussions.
This work was partly supported by the NSF grant PHY-9722060 (P.D.F.)
and by the TMR Network contract ERBFMRXCT 960012 (B.E.).

\vfill\eject

\appendix{A}{The planar free energy for small $z$}

In this appendix, we compute the leading coefficient $\varphi_1(p,q,t)$
of the free energy $f(z;p,q,t)$ when $z$ is small.  With
$\alpha^*$ the solution \sapo\ of the saddle-point equation \sappo,
we have
\eqn\jkl{\varphi_1(p,q,t)~=~S(\alpha^*)~=~p\, {\rm Log}(1+{\alpha^*\over p})
+q \, {\rm Log}(1+{\alpha^*\over q}) -\alpha^* }
Let us compute $\psi(p,q,t)\equiv t\partial_t \varphi_1(p,q,t)$ in
terms of $\alpha^*$. Using the formula \efac\ for $S(\alpha)$ and the
fact that, at the saddle point $\alpha^*$, we simply have to take 
the {\it explicit} derivative of $S(\alpha)$ wrt $t$, 
we obtain immediately
\eqn\derit{\psi(p,q,t)~=~\alpha^* }
with $\alpha^*$ given by \sapo .
Hence, introducing the reduced variables
$u=pt$ and $v=qt$, we have $\psi(p,q,t)=I(u,v)/t$,
where the function $I(u,v)$ reads
\eqn\quafuI{ I(u,v)~=~{1 \over 2}\big(
1-u-v-\sqrt{1-2(u+v)+(u-v)^2}\big) }
and satisfies the quadratic equation
\eqn\obfin{ (I+u)(I+v)-I~=~0 }
Expanding \quafuI\ in powers of $u$ and $v$, we find
\eqn\Iexp{ I(u,v)~=~
\sum_{k\geq 1} \sum_{j=1}^k I_{k,j} \, u^j \, v^{k+1-j}}
with the integer coefficients
\eqn\intco{ I_{k,j}~=~{1 \over k} {k \choose j} {k \choose j-1}}
Integrating $\psi(p,q,t)/t=I(pt,qt)/t^2$ once wrt $t$, we
finally get
\eqn\finf{ \varphi_1(p,q,t)~=~\sum_{k \geq 1} t^k \mu_k(p,q) }
where
\eqn\valmuko{\mu_k(p,q)~=~{1\over k} \sum_{j=1}^k I_{k,j}\,
p^j \, q^{k+1-j}}
and the result \valmuk\ follows.
This result can be checked directly on \finome.

From the behavior $\alpha^*\sim (t_*-t)^{1/2}$ read on \sapo ,
we deduce by integrating \derit\ that 
$\varphi_1(p,q,t)\sim (t_*-t)^{3/2}$.

\fig{Equivalence between (a) a bi-colored fatgraph with a unique 
vertex and a marked
edge, (b) a system of bi-colored arches and (c) a system of arches closed
into a set of connected circuits}{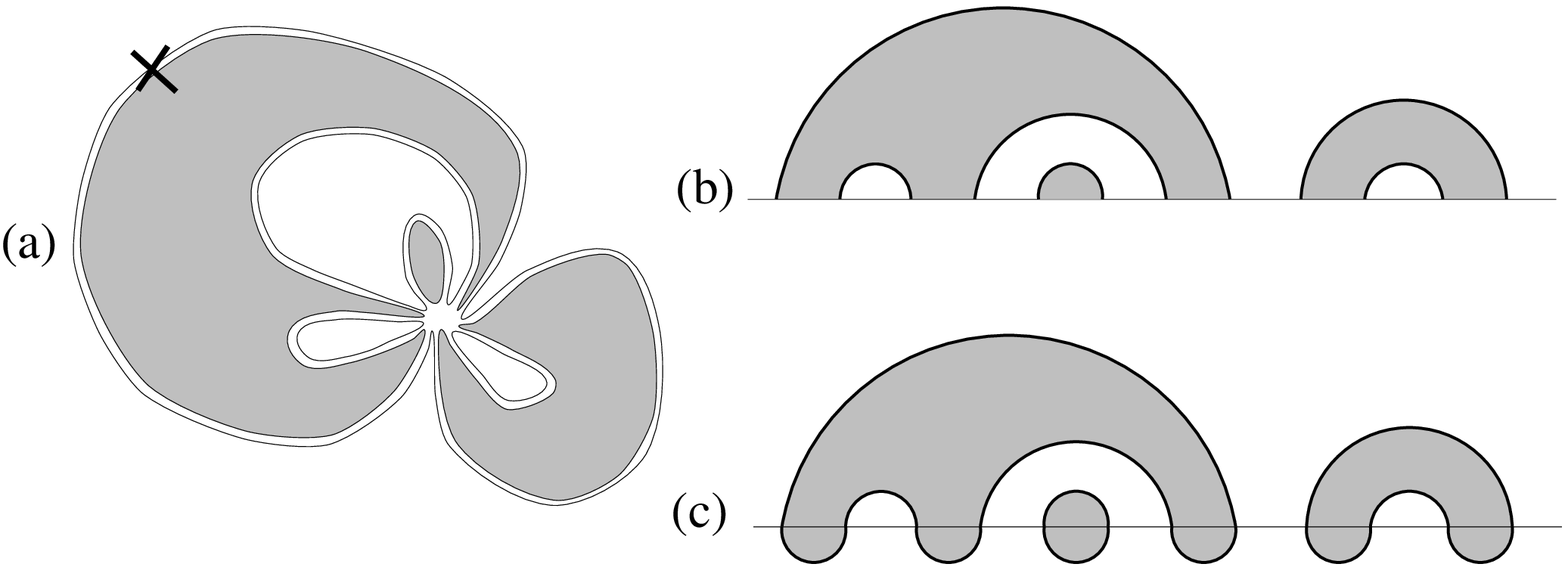}{10.truecm}\figlabel\arches 

The numbers $I_{k,j}$ \Iexp\ have already emerged in \DGG\
in the context of arch configurations.
The connection between $\psi(p,q,t)$ and arch configurations can be
stated as follows. Since $\varphi_1(p,q,t)$ keeps only the linear
term in $z$ of $f(z,p,q,t)$, it is a generating function 
for planar (dual) black and white fatgraphs with exactly 
$1$ vertex. The operation
$t\partial_t$ simply corresponds to specifying an edge among the
$E(\Gamma)$ edges of a graph $\Gamma$. We can now stretch the unique 
vertex of the graph into a straight line and let the edges of the graph 
form a system of $E(\Gamma)$ arches drawn on top of this line (see Fig.\arches ). 
The stretching operation is completely
specified if we impose that the marked edge becomes the leftmost arch, 
with its white adjacent face above it.
The function $\psi(p,q,t)$ is thus a generating function for
systems of bi-colored arches. Note that the same arch configuration 
is obtained exactly $\vert {\rm Aut}(\Gamma)\vert$ times from the different
markings of a given graph $\Gamma$, which in turn cancels the symmetry factor 
in the diagrammatic expansion.
We thus have 
\eqn\psiarches{\psi(p,q,t)~=~\sum_{k\geq 1}
\sum_{j=1}^k M_{k,j} t^k p^j q^{k+1-j}}
where $M_{k,j}$ is the number of distinct configurations of
$k$ arches drawn on top of an oriented line, separating $k+1$ domains in 
exactly $j$ black domains and $\ell=k+1-j$ 
white domains when colored in black 
and white with alternating colors, the outermost domain being white. 

The number $M_{k,j}$ has been computed in \DGG, with the
result \eqn\mkr{M_{k,j}~=~I_{k,j}} 
given in \Iexp , with a slightly different 
interpretation as the number of 
systems of $k$ arches drawn on top of a 
line which give rise to exactly $j$ 
connected circuits when closed below the line 
by elementary arches connecting 
{\it successive} points on the line in pairs. 
This number is clearly the same as
the number of bi-colored systems with $j$ black domains and $k+1-j$ 
white domains 
by coloring in black the interior of the 
$j$ connected circuits (see Fig.\arches).

The interpretation of $\psi(p,q,t)$ as the generating function 
for bi-colored systems of arches allows for a more direct 
computation by use of a simple
recursive relation for the $M_{k,j}$. Denoting by
\eqn\cpq{M_k(p,q)~=~\sum_{j=1}^k M_{k,j}\, p^j \, q^{k+1-j}}
we have the recursion relation
\eqn\reccpq{M_{k+1}(p,q)=\sum_{j=0}^k M_j(q,p) M_{k-j}(p,q)}
with the convention $M_0(p,q)=q$. This equation simply expresses
the fact that the leftmost arch separates the system of bi-colored arches
into two systems of bi-colored arches. 
Note that $p$ and $q$ are exchanged
in one of these two systems of arches.
The above recursion relation translates into a quadratic equation for
$\psi(p,q,t)=\sum_{k\geq 1}M_k(p,q)t^k$
\eqn\quadpsi{\psi(p,q,t)~=~t\big(\psi(q,p,t)+p\big)\big(\psi(p,q,t)+q
\big)}
which, since $\psi(p,q,t)$ is by definition symmetric in the exchange 
$p\leftrightarrow q$, is nothing but the equation \obfin.

Note finally that when $p=q$, $I(u,u)=u(C(u)-1)$, where
$C(u)$
is the generating function
of the Catalan numbers \catagenera\-\catalan,
and it follows that
\eqn\mupp{  \mu_k(p,p)~=~ p^{k+1}\, {c_k \over k}~=~2p^{k+1}\, 
{(2k-1)! \over k! (k+1)!} }

\appendix{B}{Existence of a critical point}

We start from the solution $t_{\pm}(u)$ of
the equation $P(u,t(u))=0$ 
\eqn\ribranapp{\eqalign{ 
pt_{\pm}(u)~&=~{u(1-u) 
\over (1-u)^2-\left({q-z\over p}\right)^2 u^2} \times \Bigg[ 
\left(1-{q+z\over p}\right)u(1-u)\cr
&\pm\sqrt{u^2(1-u)^2
\left(1-{q+z\over p}\right)^2+(1-2u)\left((1-u)^2-\left(
{q-z\over p}\right)^2u^2\right)} \Bigg]}}
We want to know whether or not the branch $t_+(u)$ 
(satisfying $pt_+(u)\sim u$ at small $u$) reaches continuously by
increasing $u$ from $0$ a maximum at some value $u_*$ between 
$0$ and $1$. We assume here that $-1\le z/p < q/p <1$.

We first note that the denominator in \ribranapp\ vanishes
for $u=u_1$ with
\eqn\uun{u_1={1\over 1+{q-z \over p}}}
between $1/3$ and $1$.  
Let us study the possible vanishing of the discriminant in
\ribranapp.
This discriminant vanishes for
\eqn\vanidis{\left(1-{q+z\over p}\right)^2=\left({1\over (1-u)^2}-
{1\over u^2}\right)\left((1-u)^2-\left({q-z\over p}\right)^2u^2\right)}
In the domain $0<u<1$, the r.h.s. is positive for $u$
between $1/2$ and $u_1$, and is maximum at 
\eqn\maxdis{u_2={1\over 1+\sqrt{q-z\over p}}} 
with the value
\eqn\maxvaldis{\left(1-{q-z\over p}\right)^2}
The discriminant will thus vanish if and only if 
\eqn\conddis{\left(1-{q+z\over p}\right)^2\le \left(1-{q-z\over p}\right)^2}
This is true in the domain $-1\le z/p < q/p <1$ if and only if $z/p\ge 0$.
In this case, the discriminant vanishes exactly once in the range
$0<u\le u_2$ for some value $u_0$
with moreover $1/2 \le u_0 \le  u_2< u_1$ (since $(q-z)/p<1$). At $u=u_0$, the
two branches $t_\pm(u)$ meet with an infinite slope (if $z/p>0$) and make a loop.
Clearly, the branch $t_+(u)$ must have a maximum at some $u_*< u_0$ and
we find a critical point. For $z/p=0$, we find a critical point 
at $u_*=u_0=u_2$.

Conversely, if $z/p<0$, the discriminant does not vanish in the range $0<u<1$
and, since $(1-(q+z)/p)>0$, we find that $t_+(u)$ is a strictly increasing
function which diverges at $u=u_1$. We find no maximum is this case, hence
no critical point. 

\fig{The behavior of $pt_\pm(u)$ for $0\le u\le 1$, here represented
for $q/p=2/3$ and decreasing values of $z/p$. A critical point corresponds
to a maximum of $pt(u)$ lying on the {\it correct} branch of solution, 
i.e. that leaving the origin with a positive slope. For $z/p>0$
(fig. a,b,c) this branch forms a loop with the {\it wrong} branch
leaving the origin with a negative slope: a maximum is reached on the correct
branch before the two branches meet. For $z/p=0$ (fig. d), a maximum is 
also reached, lying now exactly at the meeting point of the two branches. 
For $z/p<0$ (fig. e,f), the two branches avoid each other and no maximum 
is found for the correct branch.
}{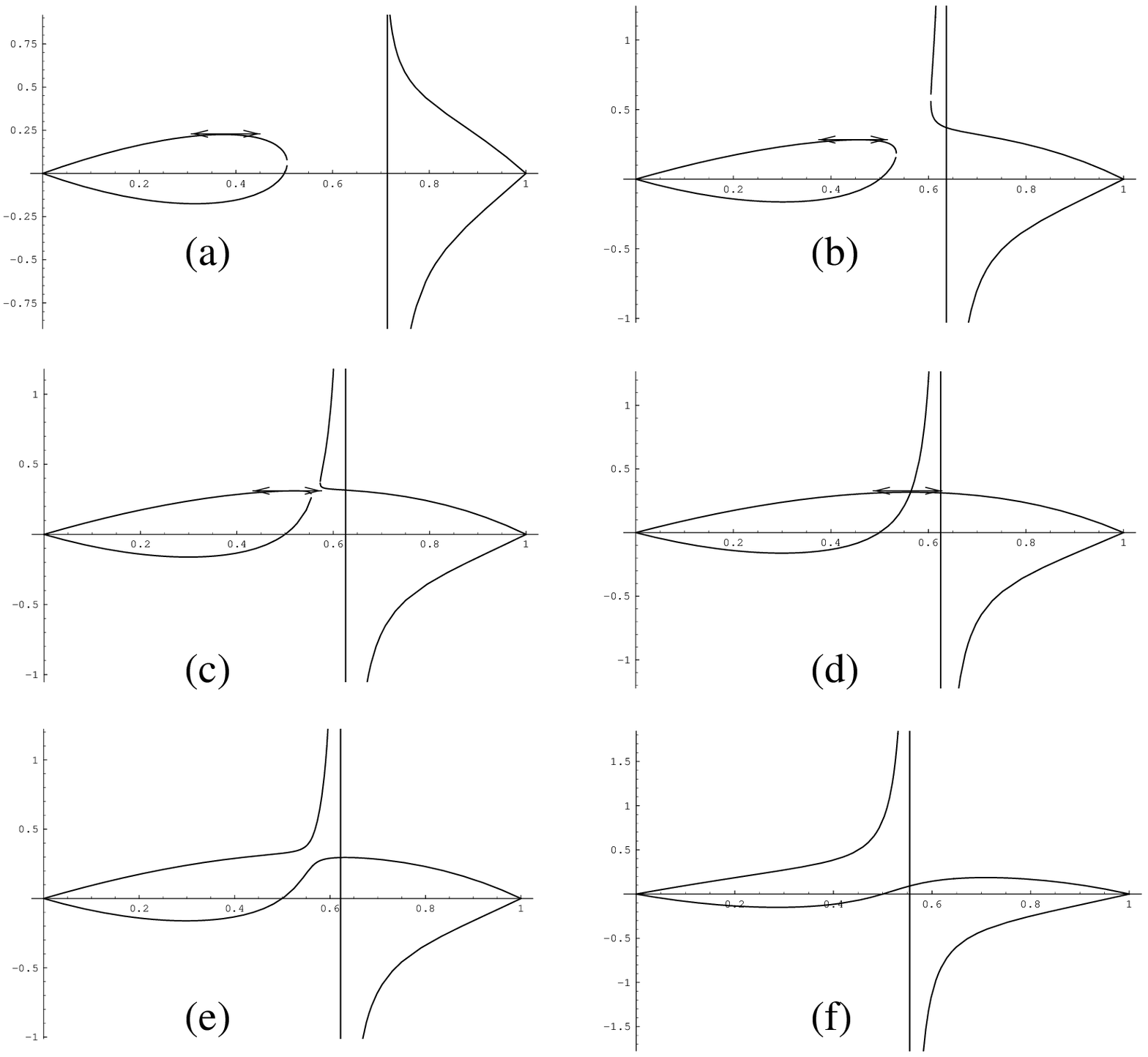}{11.cm}
\figlabel\courbes
We give an example of these behaviors in the Figure \courbes.
\listrefs
\bye